\documentclass[twocolumn,pre,preprintnumbers,superscriptaddress]{revtex4-1}
\usepackage{textcomp}
\usepackage{amsmath}
\usepackage{amssymb}
\usepackage{graphicx}
\usepackage{color}
\usepackage{soul}
\usepackage{array}

\usepackage{titlesec}
\newcommand{\be}{\begin{equation}}
\newcommand{\ee}{\end{equation}}
\newcommand{\ba}{\begin{eqnarray}}
\newcommand{\ea}{\end{eqnarray}}

\newcommand{\cmp}
{\affiliation{Condensed Matter Physics Division, 
Saha Institute of Nuclear Physics, Kolkata 700064, India}}
\newcommand{\barasat}
{\affiliation{Barasat Government College, Barasat, Kolkata 700124, India}}
\newcommand{\snb}
{\affiliation{S. N. Bose National Centre for Basic Sciences, Kolkata 700106, India}}

\begin{document}

\title{On the Question of Ergodicity in Quantum  Spin Glass Phase and its role in Quantum Annealing}

 \author{Sudip Mukherjee}
 \email{sudip.mukherjee@saha.ac.in}
 \barasat \cmp

 \author{Bikas K Chakrabarti}
 \email{bikask.chakrabarti@saha.ac.in}
 \cmp \snb

\begin{abstract}
We first review, following our earlier studies, the critical behavior of the quantum Sherrington-Kirkpatrick (SK) 
model at finite as well as at zero temperatures. Through the analysis of the Binder cumulant we determined the entire 
phase diagram of the model and from the scaling analysis of the numerical data we obtained the correlation length 
exponent. For both the critical Binder cumulant and the correlation length exponent, we observed a crossover from 
classical- to quantum-fluctuation-dominated values at a finite temperature. We studied the behavior of the order parameter 
distribution of the model in the glass phase (at finite and zero temperatures). Along with a classical-fluctuation-dominated 
nonergodic region (where the replica symmetry is broken), we also found a quantum-fluctuation-dominated low-temperature 
ergodic region in the spin glass phase. In this quantum-fluctuation-dominated region, the order parameter 
distribution has a narrow peak around its most probable value, eventually becoming a delta function in the 
infinite-system-size limit (indicating replica symmetry restoration or ergodicity in the system). We also found that the annealing 
time (to reach a very low energy level of the classical SK model) becomes practically system-size-independent when the 
annealing paths pass through this ergodic region. In contrast, when such paths pass through the nonergodic region, 
the convergence time  grows rapidly with the system size. We present a new study of the autocorrelation of the spins in 
both ergodic and nonergodic regions. We found a significant increase in the relaxation time (and also a change in the relaxation behavior) 
in the classical-fluctuation-dominated (nonergodic) region compared with that in the quantum-fluctuation-dominated (ergodic) 
region of the spin glass phase. 
\end{abstract}
\maketitle

\section{Introduction}\label{intro} 
Spin glasses~\cite{sudip-sk} have many intriguing features in their thermodynamic phases and transition behaviors.  
The effects of quantum fluctuations on such spin glass phases are being investigated extensively these days in the context
of the physics of quantum glasses and information processing. For this, we have chosen quantum Ising spin glass 
models~\cite{sudip-bkc_81,sudip-bkc-book}. We focus our study on the Sherrington-Kirkpatrick (SK) spin glass model~\cite{sudip-sk} in the presence of a transverse field~\cite{sudip-bkc-book}. Many studies have already been carried out 
(see e.g., Refs.~\cite{sudip-yamamoto,sudip-usadel,sudip-kopec,sudip-gold,sudip-lai,sudip-Takeda,sudip-Hen}) to extract 
some isolated features of the quantum phase transitions of the SK model. We have performed detailed numerical studies of the critical 
behavior of this model at finite temperatures as well as at zero temperature. We have numerically extracted the entire phase 
diagram of the model. The finite-temperature analysis was carried out by Monte Carlo simulation and the zero temperature critical behavior 
was obtained using the exact diagonalization method. From both of these numerical techniques we calculated the critical Binder cumulant~\cite{sudip-binder},  which gives the phase boundary and also  the nature of the phase transitions. We found the correlation length exponent 
from the scaling behavior of the Binder cumulant with the system size. Such studies revealed the value of the critical Binder cumulant 
and the correlation length exponent, also giving the point of crossover from their `classical' behavior (associated with the classical SK model) 
to `quantum' behavior (corresponding to that at zero temperature). Interestingly, this crossover happens at a finite 
temperature. 

Due to the random and competing spin-spin interaction, the free-energy landscape of a spin glass system is highly rugged. 
Local minima are often separated by macroscopically high free-energy barriers, which are often on the order of the system 
size (escape requiring a macroscopic fraction of spins to be reversed). This feature of the free-energy landscape induces nonergodicity in the system. The system very often becomes trapped  
at one of the local minima. As a consequence, the phenomenon of replica symmetry breaking is observed in the 
spin glass phase and the order parameter follows a broad distribution. Along with a peak at nonzero value of the order 
parameter, the distribution also contains a tail that extends up to the zero value of the order parameter. This extended tail 
does not vanish even in the thermodynamic limit. Such an order parameter distribution in the spin glass phase was suggested by 
Parisi~\cite{sudip-parisi}. 

When an SK glass is placed under a transverse field the situation becomes considerably different. In the presence of 
quantum fluctuation the system can tunnel through the high (but narrow) free-energy 
barriers~\cite{sudip-ray,sudip-das,sudip-bikas,sudip-ttc-book,sudip-troyer,sudip-Katzgraber} which essentially allows the system
to avoid becoming trapped at local free-energy minima. This phenomenon of quantum tunneling often helps the system to regain  
ergodicity and one can expect the absence of replica symmetry breaking in the spin glass phase. As a result, the order parameter 
distribution has a narrow peak around some nonzero value of the order parameter, which essentially should be a delta function 
in the thermodynamic limit~\cite{sudip-ray}. 

We numerically study the behavior of the order parameter in the spin glass phase of the quantum SK model  at both finite and zero 
temperatures. From such investigations we identify a low-temperature (high-transverse-field) ergodic region in the spin glass 
phase, where the tail of the order parameter distribution vanishes in the thermodynamic limit (indicating the convergence of 
the distribution to one with a sharp peak around the most probable value). This suggests the ergodic (or replica-symmetry-restored) 
nature of the system in this region of the spin glass phase. In the rest of the spin glass phase, we find that the tail of the order 
parameter distribution does not disappear even for an infinite system size. Thus the order parameter distribution remains the Parisi 
type~\cite{sudip-parisi} (replica-symmetry-broken, indicating nonergodicity) in this region of the spin glass phase. We also carry out  
dynamical study of the system to find the variation of the annealing time in both the ergodic and nonergodic regions. We find that 
the annealing time to reach a low-energy state from the paramagnetic phase becomes independent of system the size in the case of annealing 
down through the ergodic region. On the other hand, the annealing time grows rapidly with the system size when the 
same annealing is performed through the nonergodic region. {These discussions in the following sects.~\ref{critical_phenm} and \ref{QA_EE}
are essentially based on our earlier publications~\cite{sudip-cl_qm,sudip-op_dis}.}  

We add a new study on spin autocorrelation in the glass phase {(see sect.~\ref{corr})}. We observe that the 
relaxation behavior of autocorrelation is markedly different in the ergodic and nonergodic regions. The effective 
relaxation time of the system is much higher in the classical fluctuation dominated (nonergodic) region, whereas the 
system relaxes very quickly in the quantum-fluctuation-dominated (ergodic) region of the spin glass phase.

\begin{figure*}[htb]
 \begin{center}
 \includegraphics[width=5.5cm]{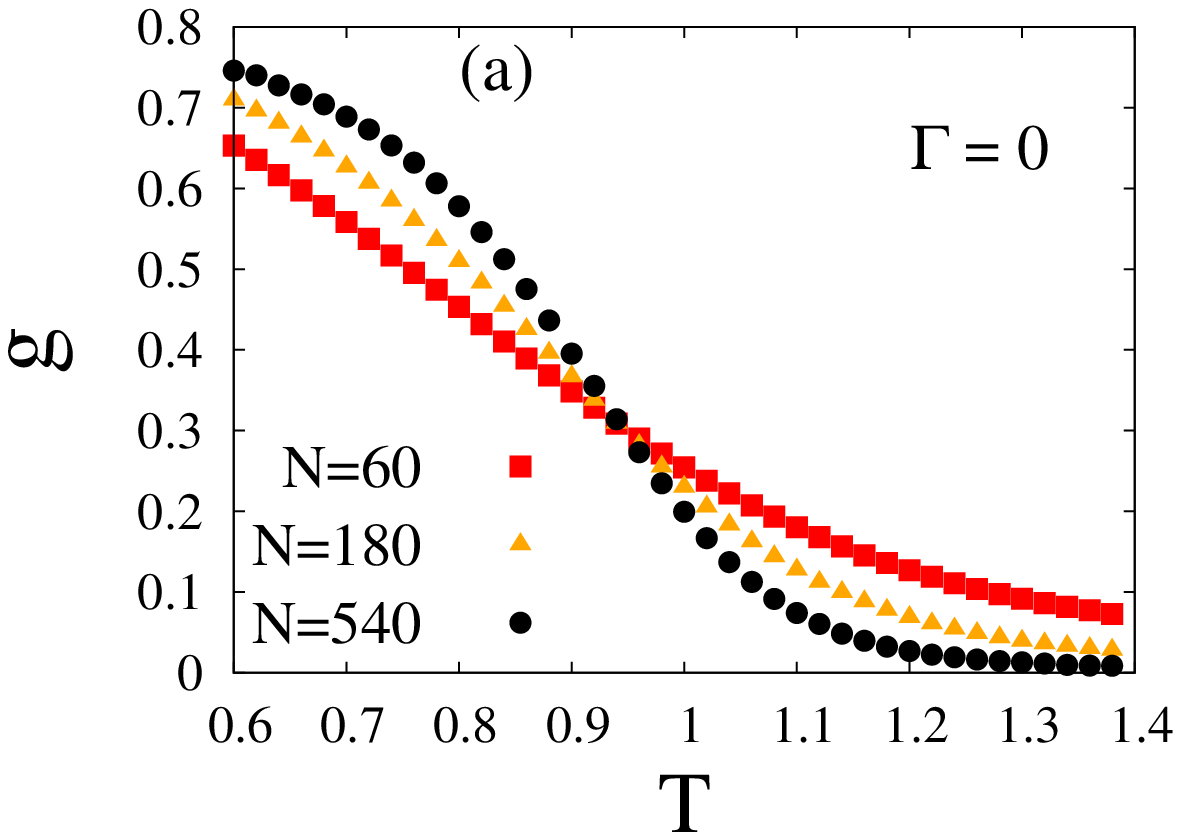}
 \includegraphics[width=5.5cm]{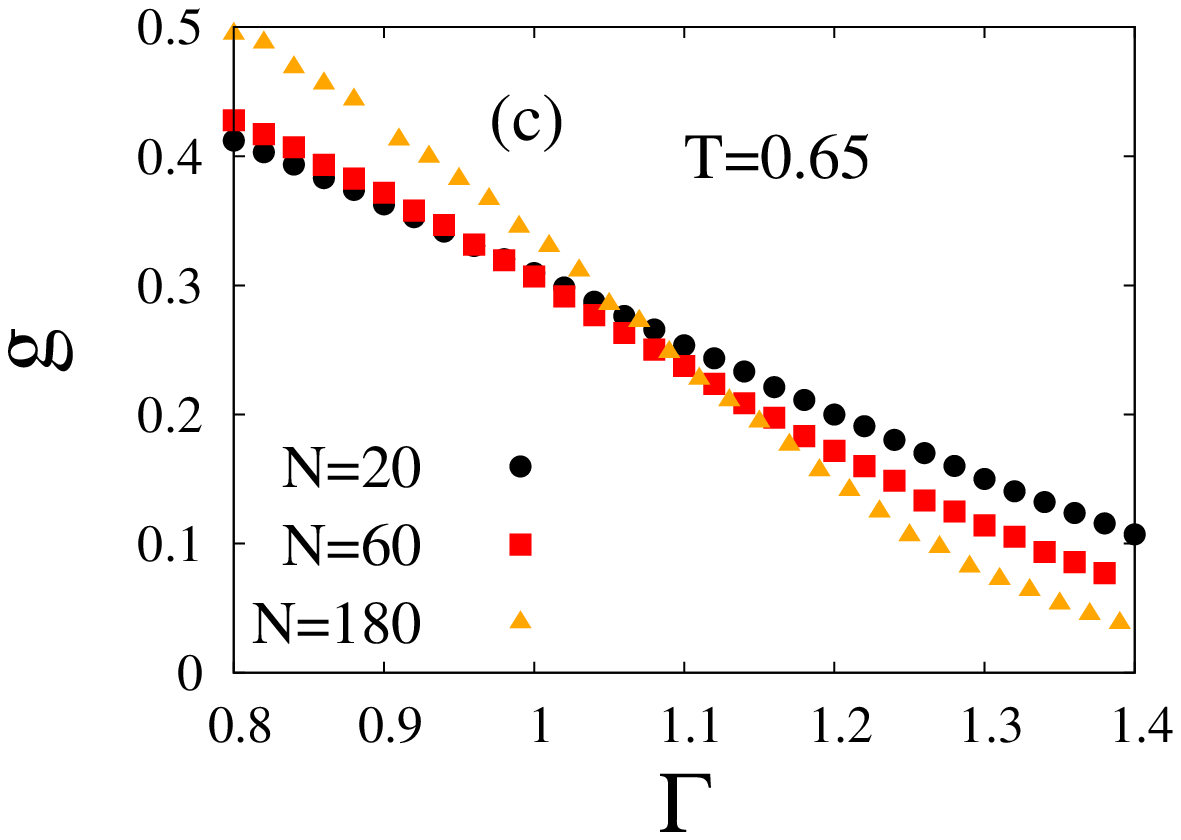}
 \includegraphics[width=5.5cm]{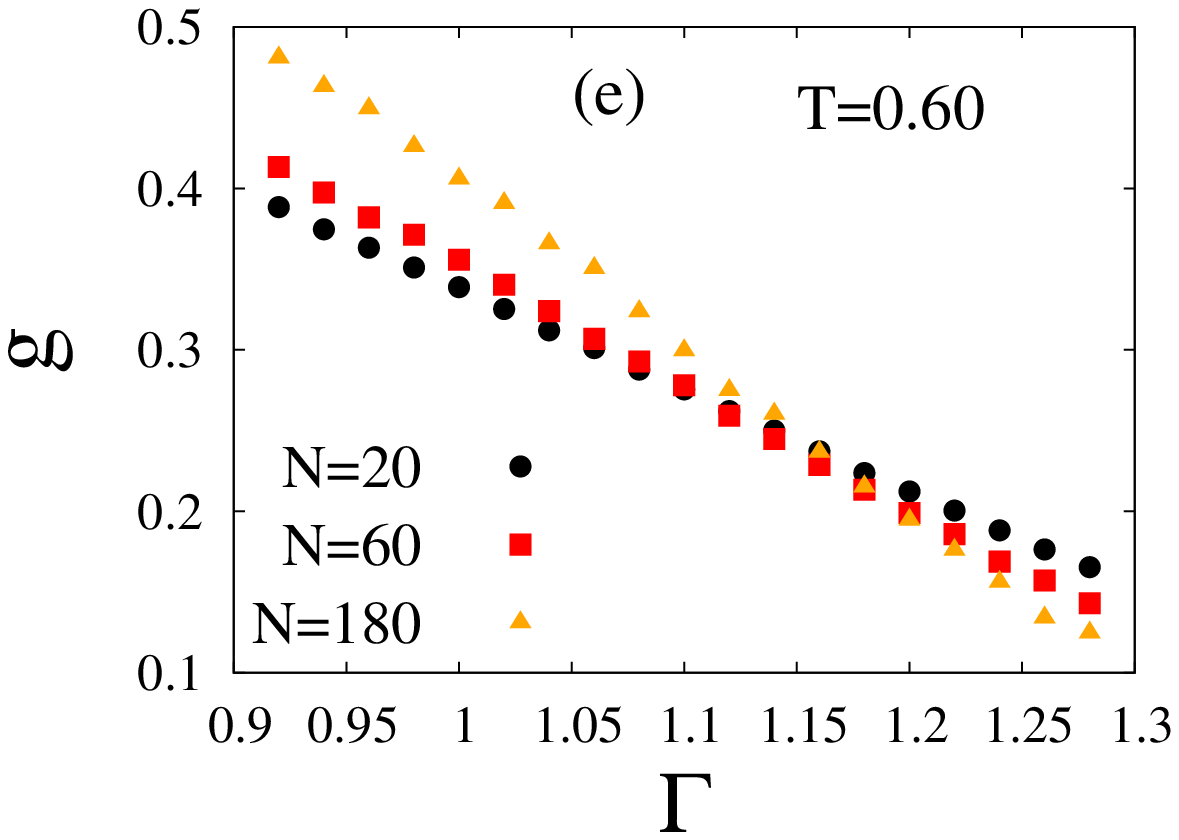}
 \includegraphics[width=5.5cm]{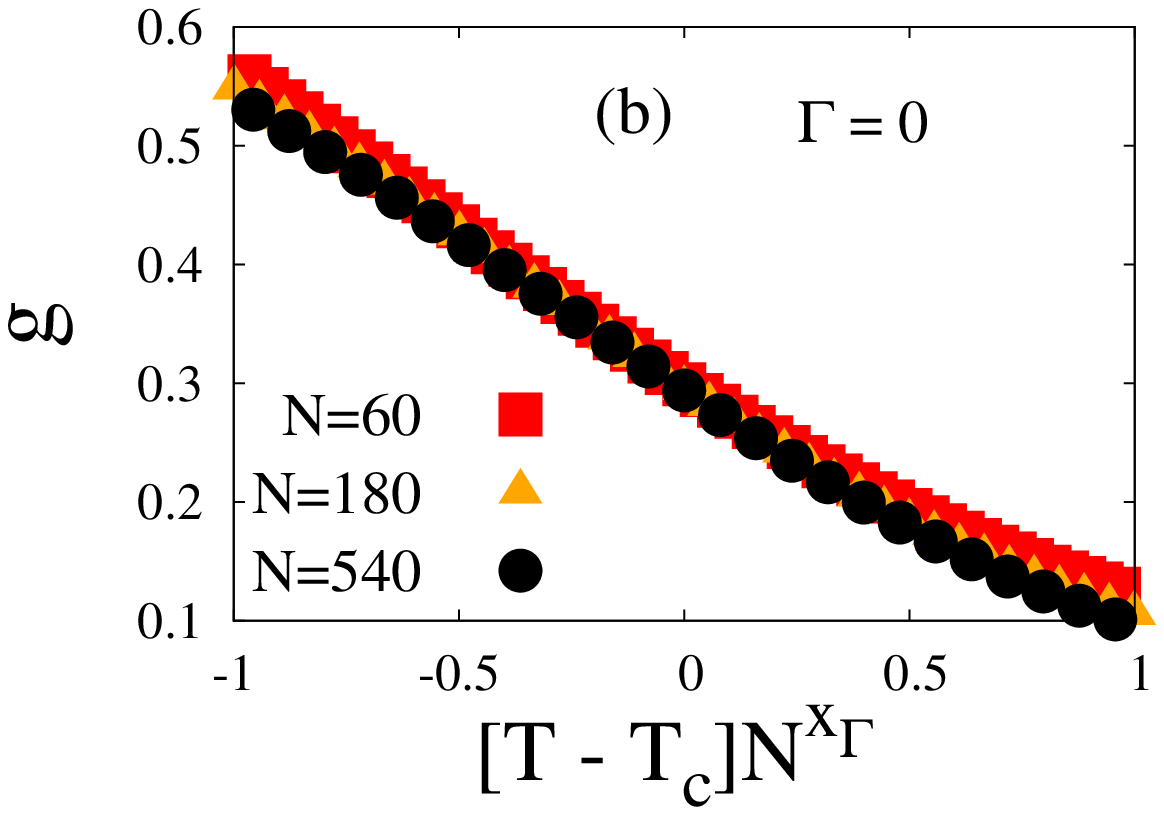}
 \includegraphics[width=5.5cm]{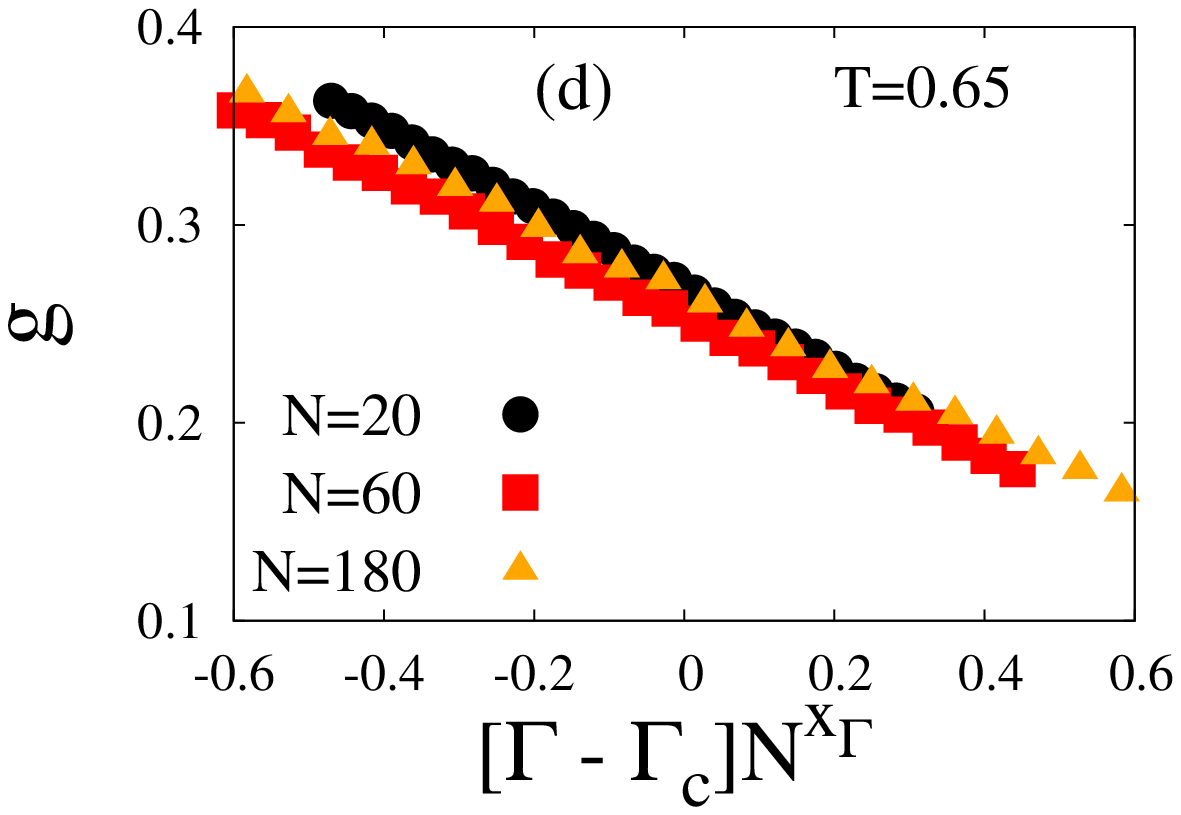}
 \includegraphics[width=5.5cm]{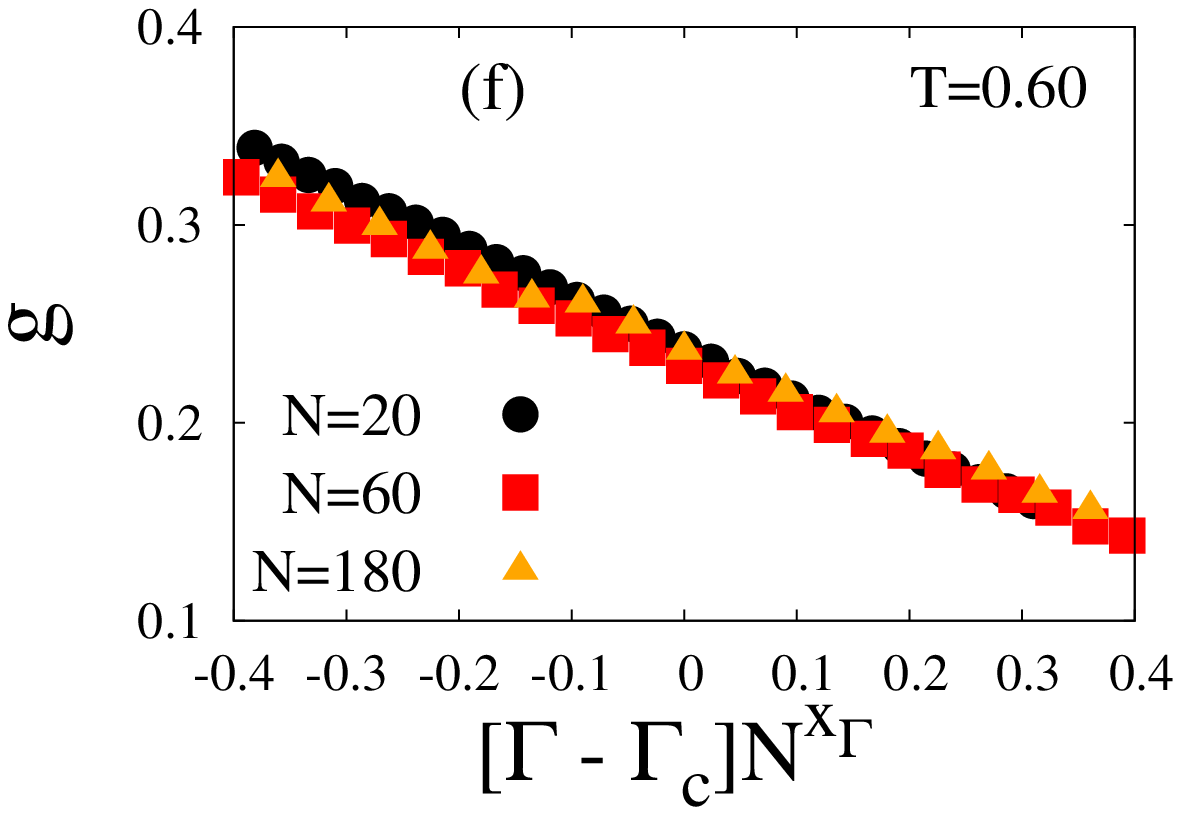}
 \end{center}
 \caption{(Color online) 
 Variation of Binder cumulant $g$ with temperature $T$ and transverse field $\Gamma$: 
 (a) for classical SK model (at $\Gamma=0$) and (c) and (e) for $T = 0.65$ and $0.60$, respectively 
 (Monte Carlo results). The interaction points give the estimate for $T_c$ or $\Gamma_c$. The symbol 
 sizes are on the order of the statistical errors of the data points. The data collapse  
 of $g$ [in  (a), (c), and (e)], when  plotted against $[T - T_c]N^{x_T}$ or $[\Gamma - \Gamma_c]N^{x_{\Gamma}}$, 
 following Eq.~(\ref{gt}), are shown in  (b), (d), and (f), respectively.  
 Such data collapses give the  values  $x_T$ or $x_{\Gamma}$ $=0.31 \pm 0.02$.}
 \label{inset_high1}
 \end{figure*}
\begin{figure*}[htb]
\begin{center}
\includegraphics[width=5.7cm]{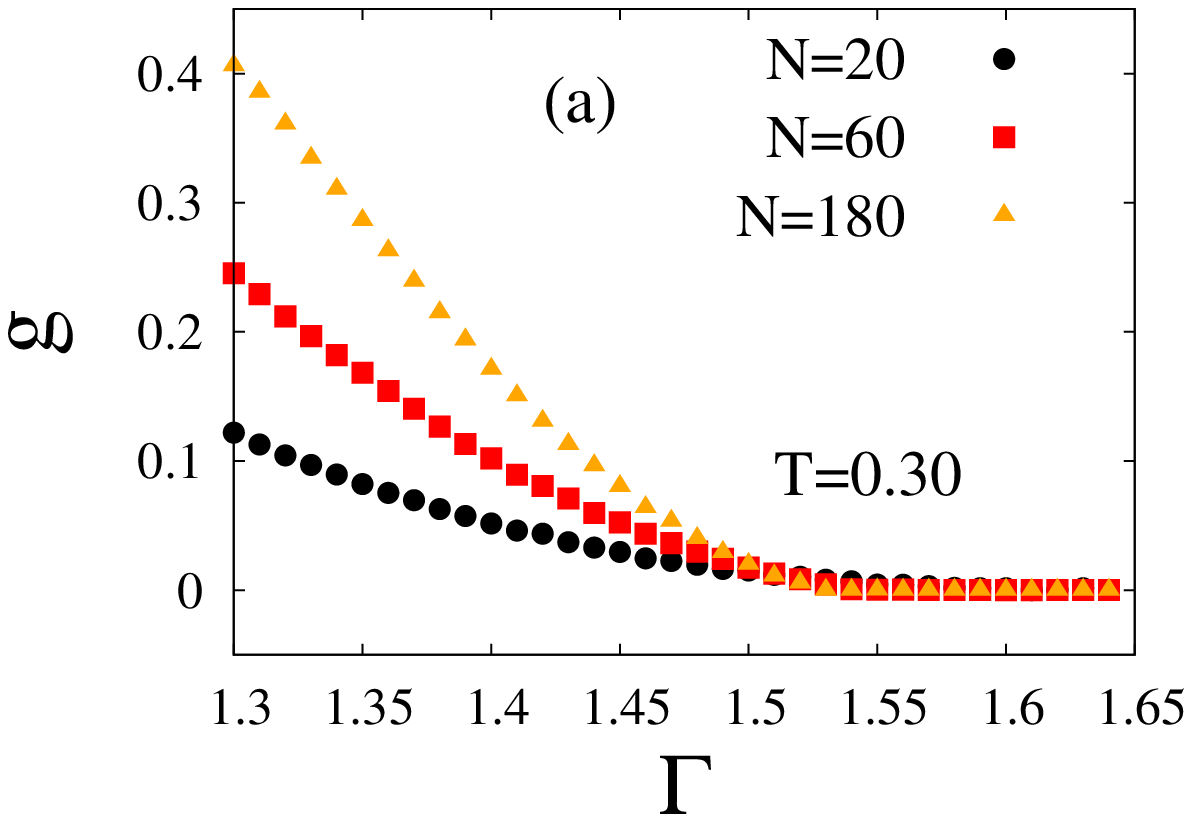}
\includegraphics[width=5.7cm]{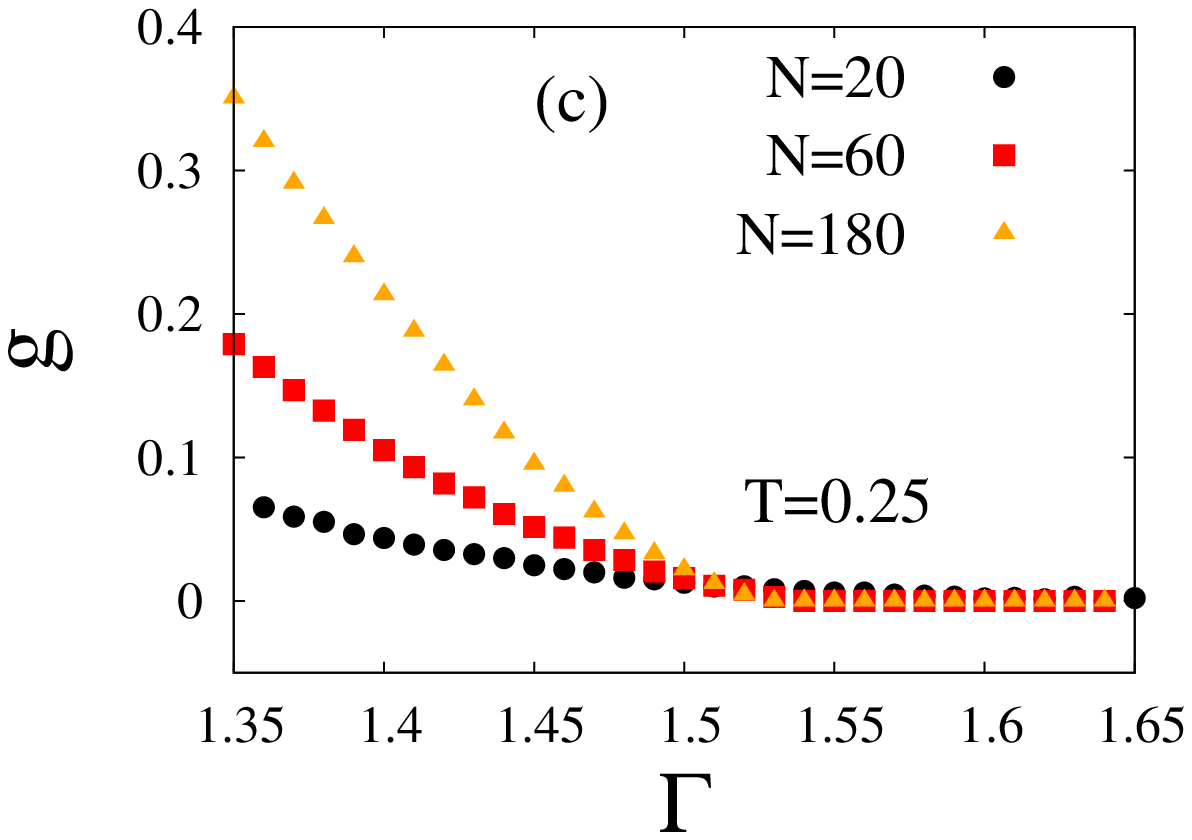}
\vskip 0.1cm
\includegraphics[width=5.7cm]{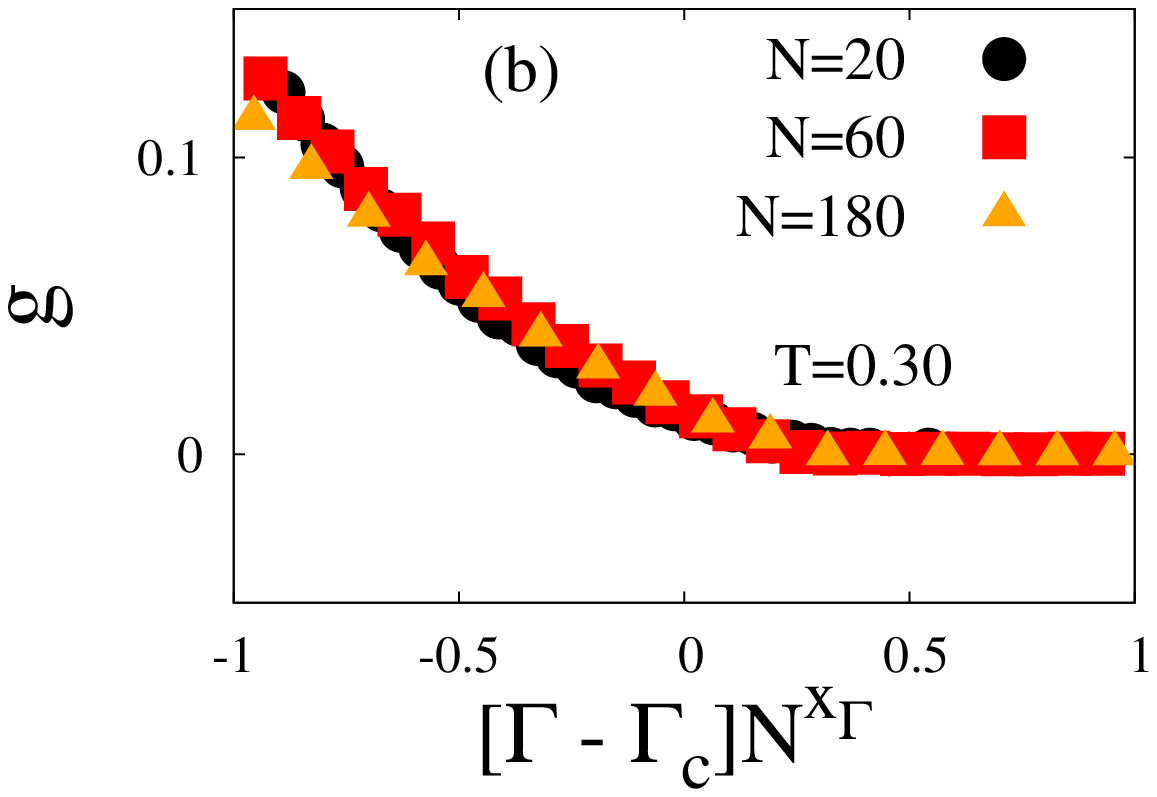}
\includegraphics[width=5.7cm]{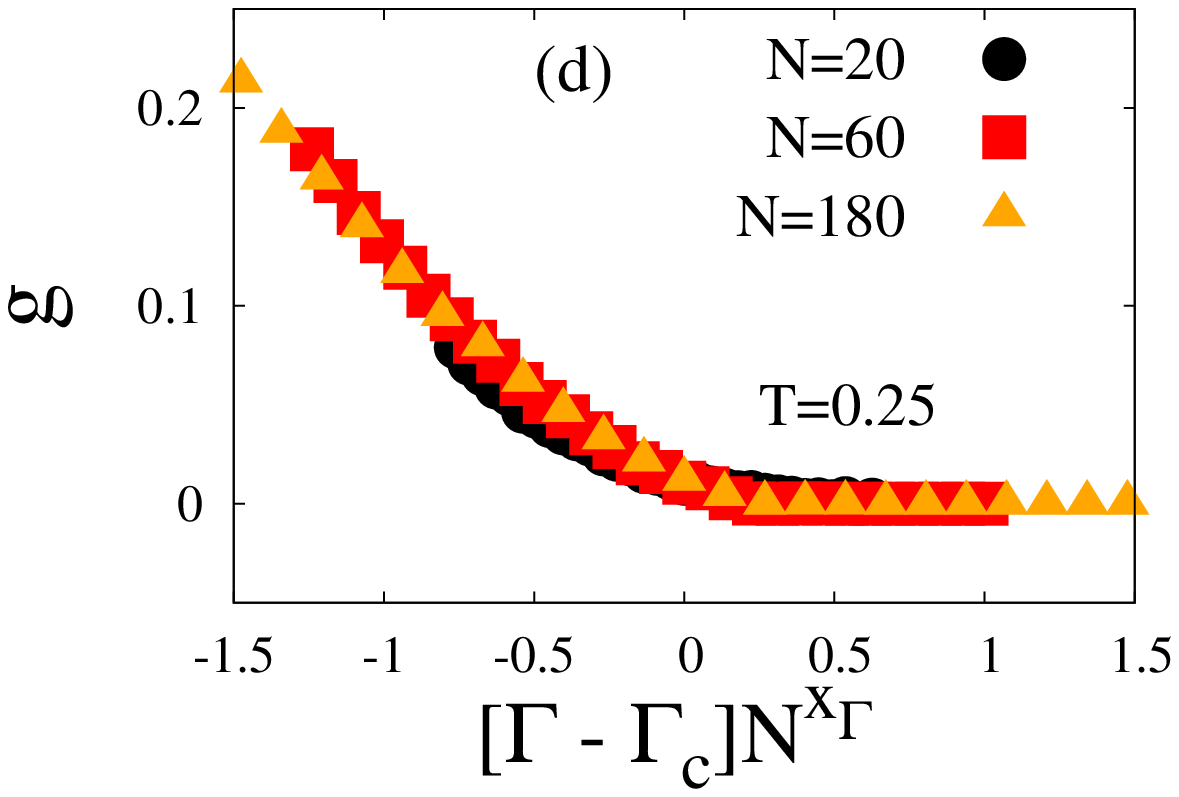}
\end{center}
\caption{(Color online)  
 Variation of Binder cumulant $g$ with transverse field $\Gamma$ at temperatures of (a) $0.30$ and (b) $0.25$ 
(Monte Carlo results). The statistical errors associated with the data points 
are on the order of the point sizes. The collapses of the $g$ curves in (a) and (c) are shown in (b) and (d), 
respectively. Again the data collapses of the $g$ curves are performed using scaling relation Eq.~(\ref{gt}) and give 
the value $x_{\Gamma}$ $=0.50 \pm 0.02$.
}
 
\label{inset_low1}
\end{figure*}
\begin{figure*}[htb]
\begin{center}
\vskip 0.1cm
\includegraphics[width=5.7cm]{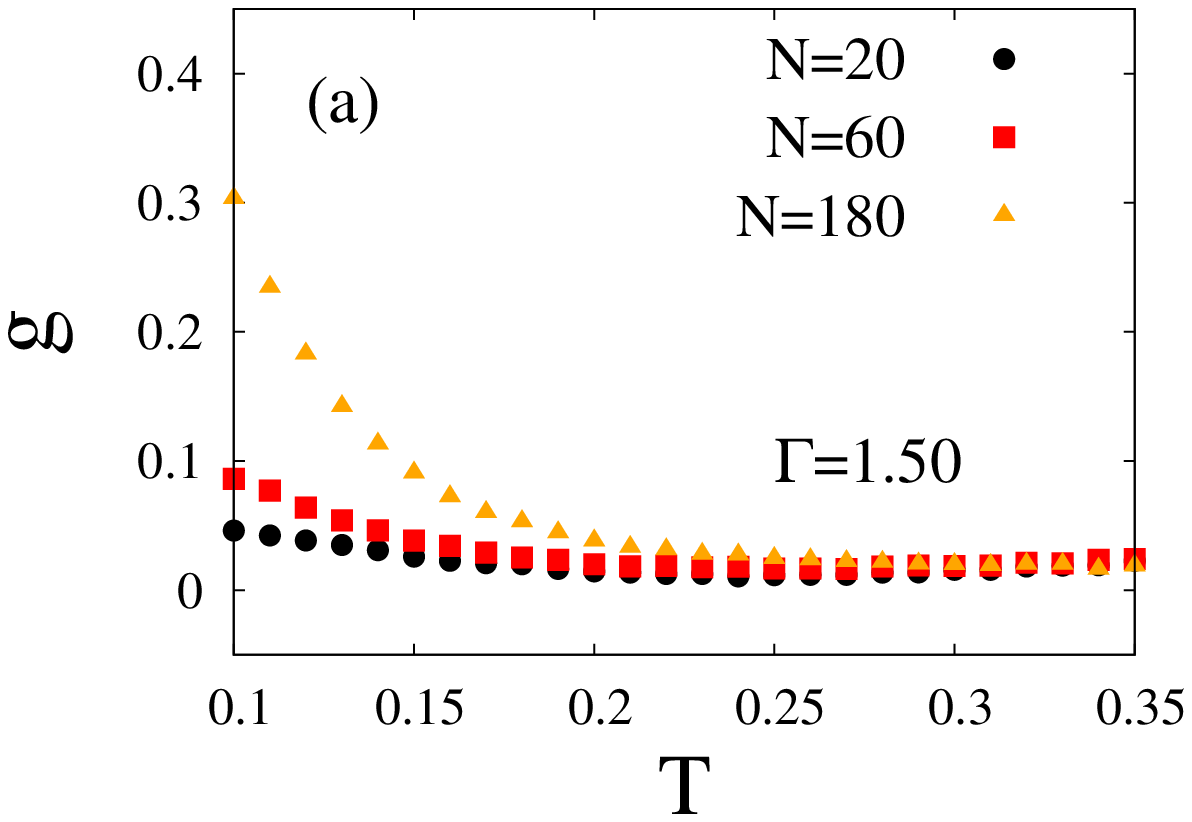}
\includegraphics[width=5.7cm]{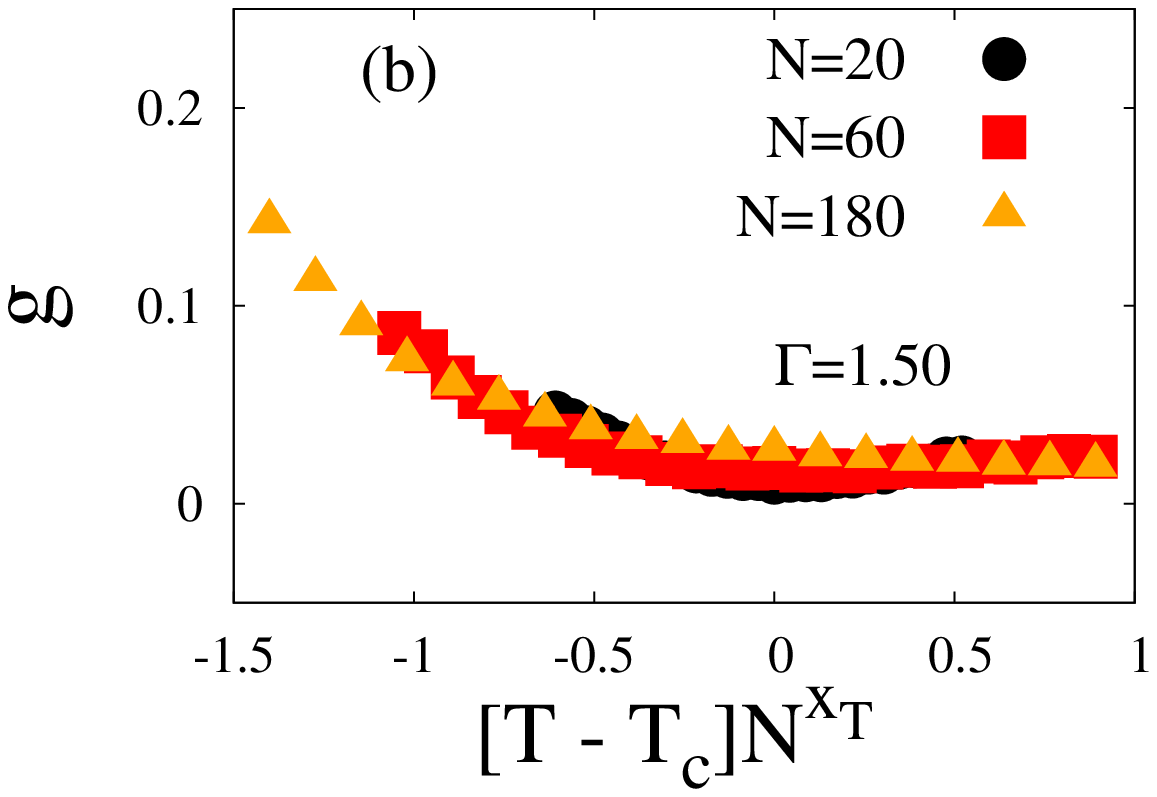}
\end{center}
\caption{(Color online) 
(a) Variation of Binder cumulant $g$ with temperature $T$ at the transverse field $\Gamma = 1.5$ 
(Monte Carlo results). (b) Data collapse of $g$ curves in (a), from which $x_T$ $=0.49$ is estimated. The 
statistical errors of the data points are on the order of the point sizes.
}
\label{inset_low2}
\end{figure*}

\section{Model}\label{model} 
The Hamiltonian of the quantum SK model with $N$ Ising spins is given by 
\begin{align}
H  = H_0 + H_I;~ H_0=-\sum_{i < j} J_{ij}\sigma_i^z\sigma_j^z;~ H_I=-{\Gamma} \sum_{i = 1}^N\sigma_i^x . \label{Ham}  
\end{align}
Here $J_{ij}$, are spin-spin interactions and they are distributed with the Gaussian distribution 
$\rho (J_{ij}) = \Big (\frac{N}{2{\pi}J^2}\Big)^{\frac{1}{2}}\exp\Big (\frac{-NJ_{ij}^2}{2J}\Big)$. The mean and 
standard deviation of the distribution are $0$ and $J/\sqrt{N}$, respectively. In this work we set $J = 1$. 
$\sigma_i^z$ and $\sigma_i^x$ are the $z$ and $x$ components of Pauli spin matrices respectively. The transverse field 
is denoted by $\Gamma$. Using the Suzuki-Trotter formalism we obtain the effective classical Hamiltonian $H_{eff}$ from 
Eq.~(\ref{Ham}) to perform Monte Carlo simulations at finite temperatures. The effective classical Hamiltonian $H_{eff}$ 
is given by 
\begin{align}
 H_{eff}=-\sum_{n=1}^M \sum_{i < j} {J_{ij}\over M}\sigma_i^n\sigma_j^n-\sum_{i=1}^N\sum_{n=1}^M{1\over {2\beta}}\text{log~coth}{\beta\Gamma\over M}\sigma_i^n\sigma_i^{n+1} \label{H_cl}. 
\end{align}
Here $\sigma_i^n=\pm 1$ is the classical Ising spin and $\beta$ is the inverse of the temperature $T$. We can see the 
appearance of an additional direction in Eq.~(\ref{H_cl}), which is often called the Trotter direction. The number of Trotter 
slices is denoted by $M$. In the limit $T\to 0$,  $M$ tends to infinity. 

\section{Study of critical behavior at finite and zero temperature}\label{critical_phenm} 
We numerically estimate the phase diagram of the quantum SK  model~\cite{sudip-cl_qm}. To find the critical transverse field 
or temperature we use the Binder cumulant technique. From the collapse of the of Binder cumulant curves for different system sizes 
we estimate the correlation length exponent. We notice a crossover in the values of the critical Binder cumulant and correlation 
length exponent at a finite temperature.  
\subsection{Monte Carlo results}
To extract the critical behavior of the quantum SK model at a finite temperature we perform Monte Carlo simulations on 
the Hamiltonian in Eq.~(\ref{H_cl}). For the study of classical SK model we simulate the Hamiltonian $H_0$. In 
each Monte Carlo step we calculate the replica overlap $q(t)=\frac{1}{NM}\sum_{i=1}^N\sum_{n=1}^M(\sigma_i^n(t))^{\phi}(\sigma_i^n(t))^{\theta}$,  
where $(\sigma_i^n)^{\phi}$ and $(\sigma_i^n)^{\theta}$ are the spins of two different replicas $\phi$ and $\theta$, respectively
corresponding to identical sets of disorder. We first allow the system to equilibrate with $t_0$ Monte Carlo steps 
then we perform thermal averaging over next $t_1$ Monte Carlo steps. We study the variation of the average Binder cumulant 
$g$ with $\Gamma$ (for fixed $T$) and $T$ (for fixed $\Gamma$) for different system sizes. In our calculation the average 
Binder cumulant is defined as~\cite{sudip-guo,sudip-alvarez} 
\begin{align}
 g=\frac{1}{2}\Big[3-\overline{\Big(\frac{\langle q^4 \rangle}{(\langle q^2 \rangle)^2}\Big)}\Big] .\label{g2}
\end{align}
Here the overhead bar indicates averaging over the configurations and $\langle.\rangle$ denotes the thermal averaging. We note 
that the average Binder cumulant can  be also defined as 
$g=\frac{1}{2}\Big[3-\frac{\overline{\langle q^4 \rangle}}{\overline{(\langle q^2 \rangle)^2}}\Big]$. With this definition of $g$ one obtains a large fluctuation and poor statistics~\cite{sudip-guo}. Thus, 
 throughout of our calculation we work with the definition of $g$ in Eq.~(\ref{g2}).

The scaling relation of $g$ near the critical region is given by $g=g(L/\xi, M/L^{z})$~\cite{sudip-guo}. Here $L$ is the 
linear size of the system and $M$ is the Trotter size. The dynamical exponent and correlation length are denoted by $z$ 
and $\xi$, respectively. The correlation length $\xi$ scales as $\xi$ $\sim (T - T_c)^{-\nu_{T}}$ or 
 $(\Gamma - \Gamma_c)^{-\nu_{\Gamma}}$ with correlation exponents $\nu_{T}$ and $\nu_{\Gamma}$. The critical temperature 
and transverse field are denoted by $T_c$ and $\Gamma_c$, respectively. Therefore the scaling relation of $g$ can be rewritten 
as 
\begin{equation}
g \sim g((T - T_c)N^{x_T},M/N^{z/d_c})~\text{or}~g((\Gamma - \Gamma_c)N^{x_{\Gamma}},M/N^{z/d_c}). \label{gt}
\end{equation}
Here $x_T= 1/\nu_Td_c$ and $x_{\Gamma}= 1/\nu_{\Gamma}d_c$. We correlate the linear dimension $L$ with the total number of 
spins $N$ through the relation $L=N^{1/d_c}$, where $d_c$ is the effective dimension of the system. We estimate the values 
of the critical transverse field $\Gamma_c$ and critical Binder cumulant $g_c$ from the intersection of the $g$ versus $\Gamma$ 
curves for different system sizes (keeping $M/L^{z}$ fixed). Using the scaling relation in Eq.~(\ref{gt}) we collapse the 
$g$ curves and estimate the values of $x_{\Gamma}$ and $x_T$.

We simulate the Hamiltonian in Eq.~(\ref{H_cl}) with system sizes $N = 20, 60, 180$. We start with $M = 10$ for the system size $N = 20$, 
and to keep $M/L^{z}$ fixed we take $M = 21, 43$ for the system sizes $N = 60, 180$, respectively. Here we consider $d_c = 6$ 
and $z = 4$~\cite{sudip-Billoire}, which are associated with the classical SK model. As there is no additional Trotter dimension in 
the Hamiltonian $H_0$, we are able to perform Monte Carlo simulations of the classical SK model with larger system sizes $N = 60, 180, 540$.
We take $t_0 = 75000$ Monte Carlo steps for the equilibration of the system and the thermal average is taken the over the next $25000$ 
 Monte Carlo steps. The disorder averaging is carried out over $1000$ samples. We observe that in the range starting from the classical 
SK model at $\Gamma = 0$ to almost $T \simeq 0.50$ ($\Gamma \simeq 1.30$), the value of $g_c$ stays almost constant at $0.22 \pm 0.02$ 
[see  Figs.~\ref{inset_high1}(a), \ref{inset_high1}(c), and \ref{inset_high1}(e)]. We also find a satisfactory data collapse of $g$ curves with 
$x_T=x_{\Gamma}= 0.31\pm 0.02$ [see Figs.~\ref{inset_high1}(b), \ref{inset_high1}(d), and \ref{inset_high1}(f)]. 
We find that the value of $g_c$ becomes vanishingly small in the range $T=0.30$ ($\Gamma \simeq 1.50$)
to $T = 0.20$ ($\Gamma \simeq 1.54$), but in this case we are unable to collapse the $g$ curves for any of the chosen values of $x_{\Gamma}$. 
In this range we repeat our simulation with $d_c=8$ and $z=2$, which are values related to the quantum SK model~\cite{sudip-david,sudip-read}.
In order to keep $M/L^{z}$ constant with these new values of $d_c$ and $z$, we take Trotter sizes $M=10, 13, 17$ for the system sizes 
$N=20, 60, 180$, respectively. We again notice that the value of $g_c$ becomes almost zero [see Figs.~\ref{inset_low1}(a) and \ref{inset_low1}(c)] 
and this time we obtains a satisfactory data collapse of $g$ curves [see Figs.~\ref{inset_low1}(b) and \ref{inset_low1}(d)] with 
$x_{\Gamma} = 0.50 \pm 0.02$. Note that, with the quantum values of $d_c$ and $z$ we are unable to collapse the $g$ 
curves consistently in the range ($\Gamma=0$, $T\simeq1.0$) to ($\Gamma \simeq 1.30$, $T\simeq0.50$). Therefore, we find a change 
in the values of $x_{\Gamma}$ and $g_c$ at low temperatures. To confirm this observation we investigate the variation of $g$ with 
$T$ for a fixed value of $\Gamma$. This variation for $\Gamma = 1.5$ is shown in Fig.~\ref{inset_low2}(a) and the corresponding data 
collapse with $x_T = 0.49$ is shown in Fig.~\ref{inset_low2}(b). This implies that at low temperatures (high $\Gamma$) the  
critical exponents are $x_T \simeq x_{\Gamma} \simeq 0.50$. The crossover in the values of $g_c$ and $x_{\Gamma}$ ($=x_T$) with 
the $\Gamma$ (or $T$) values within this range ($0.5<T<0.35$, $1.30<\Gamma<1.45$) may be abrupt. From our numerical studies here it 
is not possible to state firmly whether this crossover is gradual or abrupt.    

\begin{figure}[ht]
\begin{center}
\includegraphics[width=5.5cm]{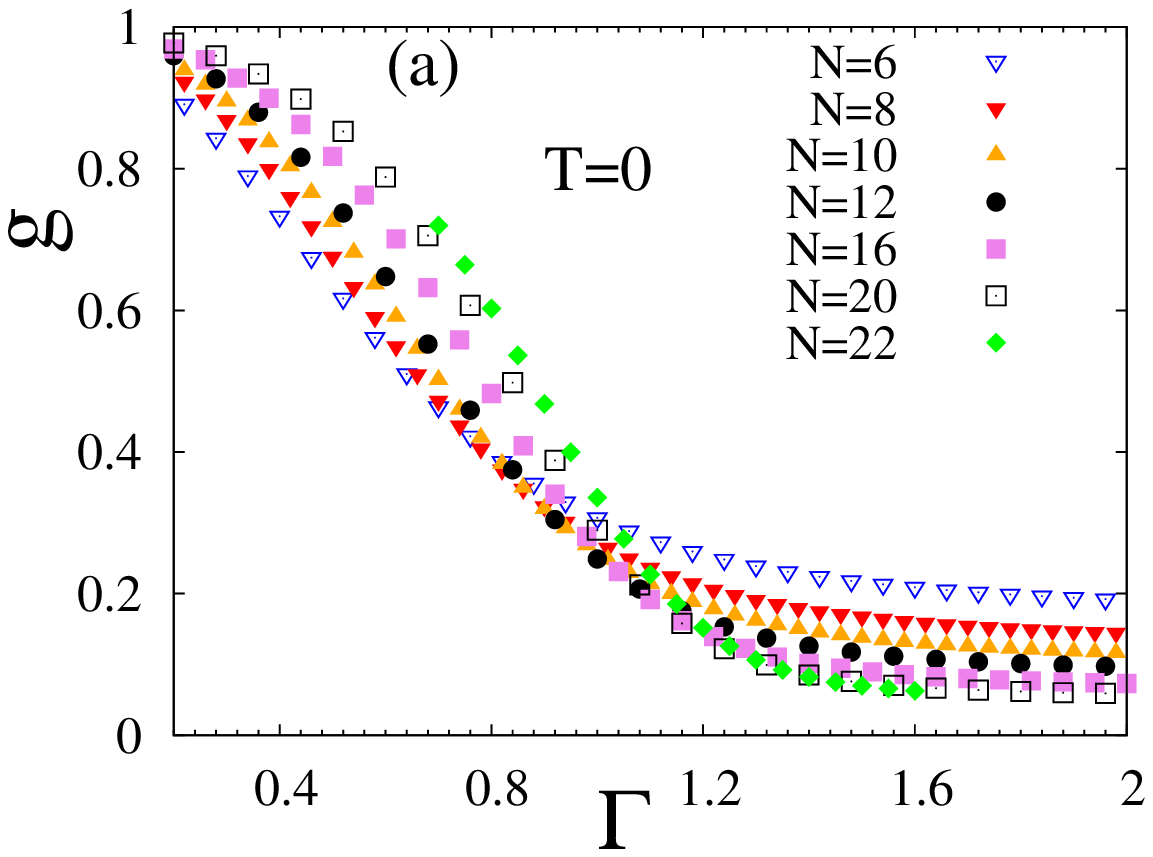}
\includegraphics[width=5.5cm]{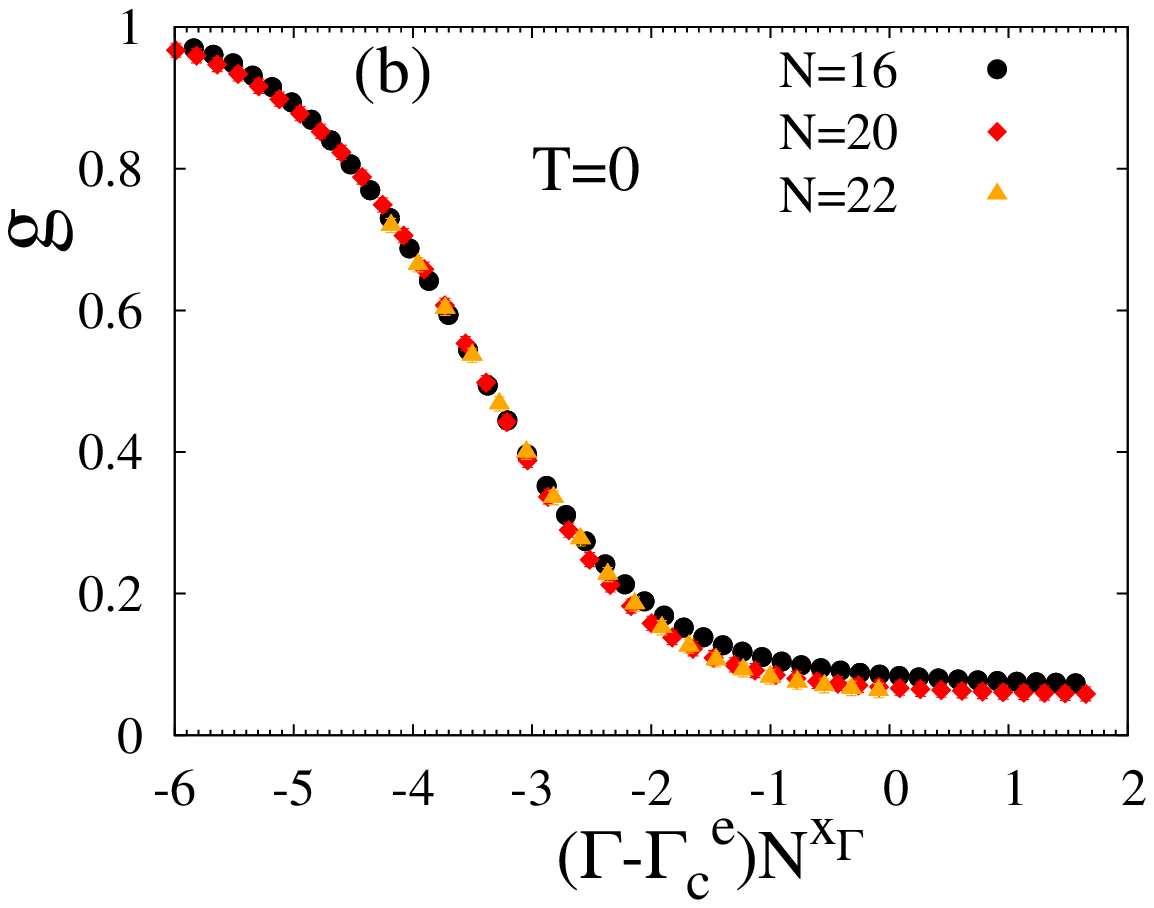}
\end{center}
\caption{(Color online)   
 (a) Exact diagonalization results of the Binder cumulant $g$ plotted as a function of $\Gamma$ at 
$T = 0$ (quantum SK model) for different system sizes. The results for the larger system sizes intersect 
at larger values of $\Gamma$, indicating the finite-size effect of the system. (b) Data collapse of 
the Binder cumulant curves for different system sizes following the scaling relation in Eq.~(\ref{gt}) with $M=0$.  
The estimated values of  $\Gamma_c^e$ and the exponent $x_{\Gamma}$ are $1.63$ and $0.5$, respectively.}
\label{bc_gama}
\end{figure}
\begin{figure}[ht]
\begin{center}
\includegraphics[width=9.5cm]{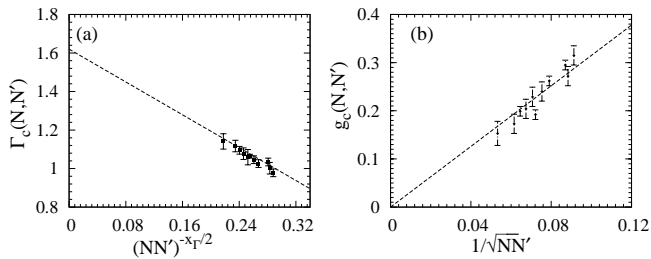}
\end{center}
\caption{   
(a) Extrapolation of critical transverse field ($\Gamma_c(N,N')$) as a function of $(NN')^{-x_{\Gamma}/2}$, 
where $N$ and $N'$ are two different system sizes. The extrapolated value of $\Gamma_c$ is $1.62$. (b) Extrapolation 
of critical Binder cumulant as a function of $1/\sqrt{NN'}$. The value of $g_c$ tends to zero in the infinite-system-size limit. For both 
 plots the best-fit lines are also shown in the figures.}
\label{gc_bc}
\end{figure}

\subsection{Zero-temperature diagonalization results}
We explore the zero-temperature critical behavior of the SK model through Binder cumulant analysis using the exact diagonalization 
technique. The diagonalization of the quantum spin glass is performed using the Lanczos algorithm. In the zero-temperature 
analysis we are able to work with the system sizes only up to $N = 22$. We construct the Hamiltonian Eq.~(\ref{Ham}) in the spin basis, 
which is made up of the eigenstates of $\sigma_i^{z} (i = 1, 2, 3,..., N)$. The $n$th eigenstate of $H$ can be expressed as 
$|\psi_n\rangle~= \sum_{\alpha=0}^{2^{N-1}} a_{\alpha}^n |\varphi_\alpha\rangle$. Here $|\varphi_\alpha\rangle$ are the eigenstates 
of $H_0$ with expansion coefficients $a_{\alpha}^n=\langle\varphi_{\alpha}|\psi_n\rangle$. For the zero-temperature analysis we define 
the order parameter as $Q = (1/N) \sum_i \overline{\langle\psi_0|\sigma_i^z|\psi_0\rangle^2}$. Since our interest is focused on 
zero-temperature analysis, we are confined to ground state ($|\psi_0\rangle$) averaging in the evaluation of the order parameter and other 
physical quantities. In this case the configuration average is again indicated by the overhead bar. The various moments of the order 
parameter can be calculated using the relation~\cite{sudip-sk,sen_97}, 
\be
 Q_k = {1\over {N^k}} \sum^{N}_{i_1} \ldots \sum^{N}_{i_k}\langle\psi_0|\sigma_{i_1}^z \ldots \sigma_{i_k}^z|\psi_0\rangle^2.
\label{moment}
\ee
Physically the $Q_k$ are the $k$-spin correlation functions for a given disorder configuration. In the case of zero-temperature, using 
Eq.~(\ref{moment}) we can define the Binder cumulant as $g=\frac{1}{2}\Big[3-\overline{\Big(\frac{ Q_4 }{( Q_2)^2}\Big)}\Big]$.

The variations of $g$ as a function of $\Gamma$ (at $T = 0$) for different system sizes are shown
in Fig.~\ref{bc_gama}(a). The finite-size effects in the estimations of $g_c$ and $\Gamma_c$ are 
quite evident due to the noncoincidence of the intersection points of the $g$ curves associated with  
the different system sizes. To account for this finite-size effect we evaluate the values of $g_c(N,N')$ 
and $\Gamma_c(N,N')$ from the intersection of the $g$ vs $\Gamma$ curves for the two system sizes 
$N$ and $N'$. Accounting for all possible pairs, we extrapolate $\Gamma_c(N,N')$ as a function of $(NN')^{-x_{\Gamma}/2}$ 
to find $\Gamma_c$ in the thermodynamic limit. Due to the absence of any known finite-size scaling behavior
of $g$, we extrapolate $g_c(N,N')$ as a function of $1/\sqrt{NN'}$ to obtain the critical Binder cumulant value for 
an infinite system size. The best fitting of $\Gamma_c(N,N')$ against $(NN')^{-x_{\Gamma}/2}$ is obtained 
for $x_{\Gamma} = 0.51$, and the extrapolated value of $\Gamma_c(N,N')$ is $1.62~\pm~0.03$ [see Fig.~\ref{gc_bc}(a)]. 
Considering the estimated critical transverse field $\Gamma_c^e=1.62$ and $x_{\Gamma} = 0.51$, we
also obtain a satisfactory data collapse of the $g$ curves associated with the different system sizes [see Fig.~\ref{bc_gama}(b)].  
From the extrapolation of $g_c$ we find that in the limit  $N,N'\rightarrow\infty$ the value of $g_c$ 
becomes very near to zero [see Fig.~\ref{gc_bc}(b)]. This observation is consistent with the Monte Carlo 
results at low temperatures. Thus, we can conclude that starting from around $T=0.35$ to $T=0$ the values of
$g_c$ as well as $x_{\Gamma}$ remain constant at $g_c \simeq 0$ and $x_{\Gamma} \simeq 0.50$.

\begin{figure}
\begin{center}
\includegraphics[width=6.2cm]{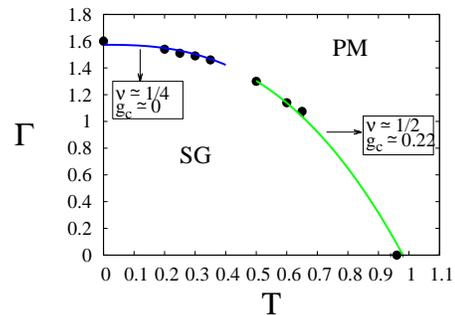}
\end{center}
\caption{(Color online) 
Phase diagram of quantum SK spin glass model as estimated from 
the Monte Carlo simulation and exact diagonalization discussed in this sect.~\ref{critical_phenm}. 
The point sizes are on the order of the statistical 
errors of the corresponding data. The spin glass and paramagnetic phases are denoted by SG and PM respectively. 
The data points at $T = 0$ and $\Gamma = 0$ are associated with the purely quantum and classical phase transitions, respectively.  
The two observed different critical behaviors of the system are indicated by {blue} ($g_c \simeq 0$, $\nu \simeq 1/4$) 
and {green} ($g_c \simeq 0.22$, $\nu \simeq 1/2$) lines. The crossover in the critical behavior occurs at around 
$T \simeq 0.49$ and $\Gamma \simeq 1.31$.}
\label{phase_diagram}
\end{figure}

\subsection{Phase diagram}
From the numerical results of the Monte Carlo simulations and the exact diagonalization related to the calculation 
of the Binder cumulant, we estimate the entire phase diagram (see Fig.~\ref{phase_diagram}) of the quantum SK spin 
glass. From the exploration of this phase diagram, we find that the value of $g_c$ remains fairly constant 
at $0.22 \pm 0.02$ in the range $T \simeq 1.0$ ($\Gamma = 0$) to $T \simeq 0.49$ ($\Gamma \simeq 1.31$). In this  
range the phase transitions are dominated by classical fluctuation (high $T$ and low $\Gamma$). On the other hand, 
beyond the point ($T \simeq 0.49, \Gamma \simeq 1.33$) to the quantum transition point ($T = 0$, $\Gamma \simeq 1.63$) 
the critical Binder cumulant $g_c$ assumes a very low value ($<0.03$) and the phase transitions are predominantly 
governed by quantum fluctuation (at low $T$ and high $\Gamma$). The two values of $g_c$ indicate two distinct 
universality classes in the critical behavior of the SK spin glass. To confirm the existence of two different universality 
classes we calculate the correlation length exponent $\nu$ on the two parts of the phase boundary associated with the two 
different values of $g_c$. In the case of classical fluctuation dominated phase transitions, if we consider 
$d_c=6$~\cite{sudip-Billoire} and $x_T = x_{\Gamma} = 1/3$, then using the relation $x_{\Gamma}= x_T=1/d_c\nu$ 
we find that $\nu = 1/2$. This value of $\nu$ is consistent with the earlier estimation of the correlation length exponent 
of the classical SK model~\cite{sudip-Billoire}. Similarly, for quantum fluctuation dominated transitions with 
$d_c=8$~\cite{sudip-david,sudip-read} and  $x_{\Gamma} = 1/2$ we obtain $\nu = 1/4$, which again shows good agreement 
with the earlier estimates~\cite{sudip-david,sudip-read}. Such changes in the values of $g_c$ and $\nu$ clearly indicate 
a finite temperature crossover between classical and quantum fluctuation dominated critical behaviors in an SK spin glass. 

\section{Study of order distribution at finite and zero temperature}\label{op_distb}
To probe the issue of ergodicity in the spin glass phase we investigate the nature of the order parameter distribution at both finite 
and zero temperatures~\cite{sudip-op_dis}. Such study clearly indicates two distinct behaviors of the order parameter distribution 
in two different regions of the spin glass phase, from which we are able to identify the ergodic and nonergodic regions in the spin 
glass phase. 

\begin{figure*}[htb]
\begin{center}
\includegraphics[width=5.5cm]{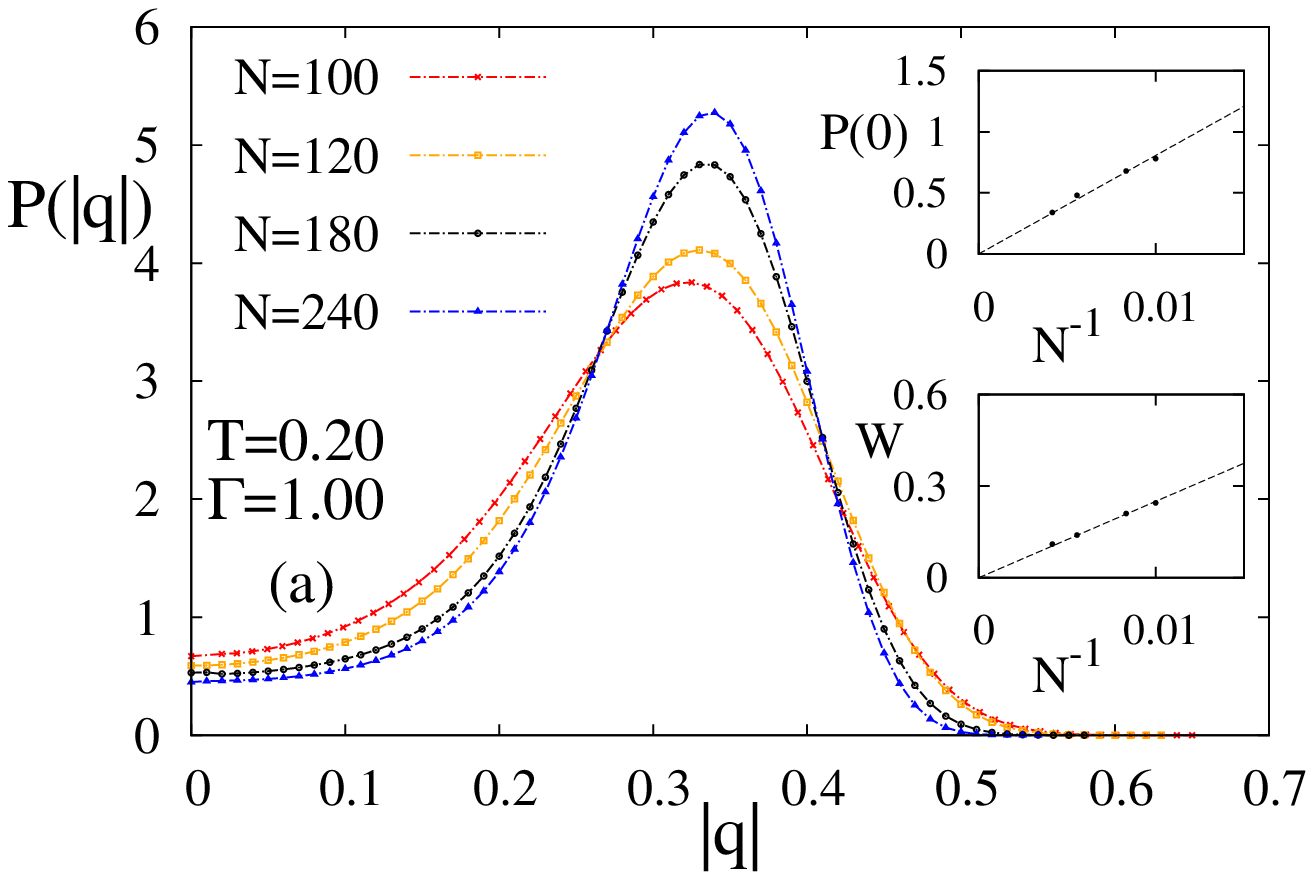}
\includegraphics[width=5.5cm]{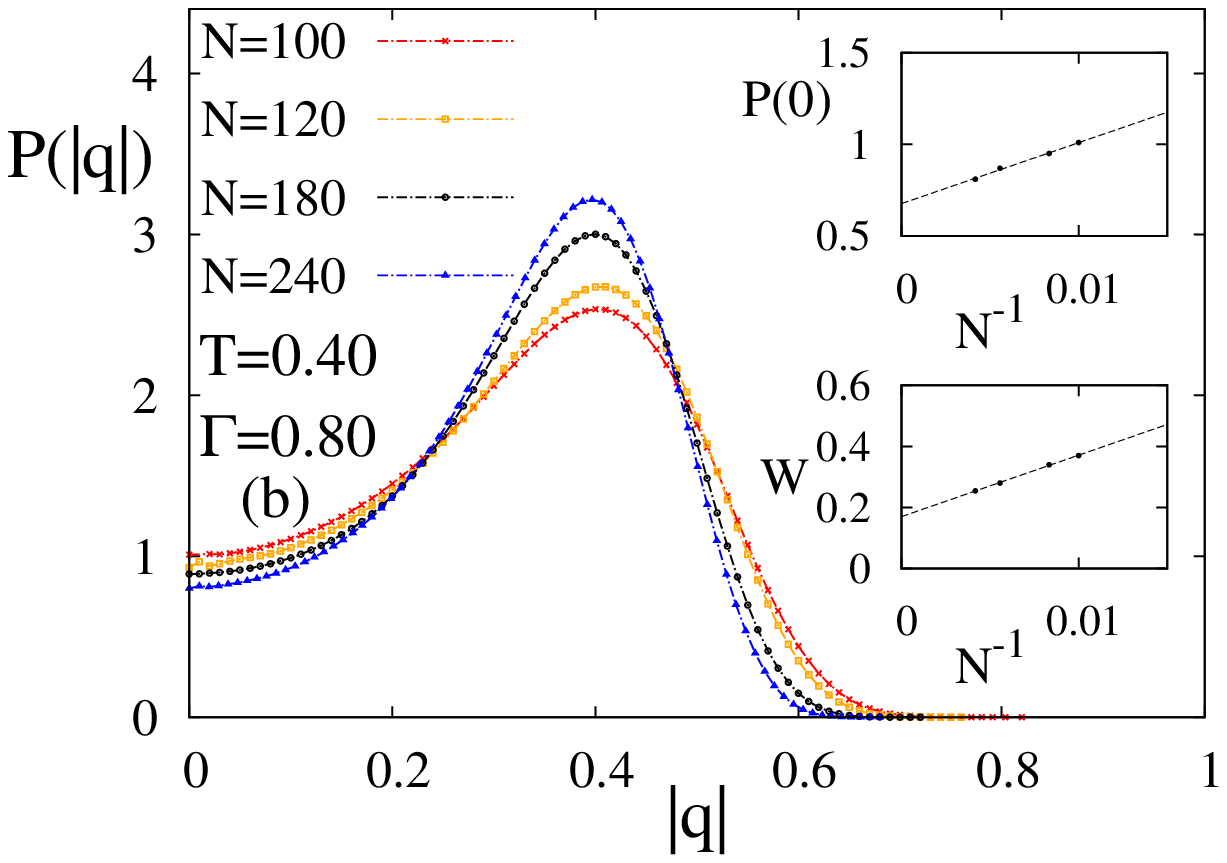}
\end{center}
\caption{(Color online) 
Plots of the area-normalized order parameter distribution $P(|q|)$ for given sets of transverse field 
$\Gamma$ and temperature $T$, obtained from Monte Carlo simulations: (a) for $T=0.20$ and $\Gamma = 1.00$, 
(b) for $T = 0.40$ and $\Gamma = 0.80$. The insets show the extrapolations of $P(0)$ and $W$ as functions of $1/N$. In the first 
case the extrapolated values of both $P(0)$ and $W$ tend to zero for an infinite system size,  whereas in the other case 
the values of these quantities do not vanish even in the thermodynamic limit.}
\label{area_ergodic}
\end{figure*}
\begin{figure*}[htb]
\begin{center}
\includegraphics[width=5.5cm]{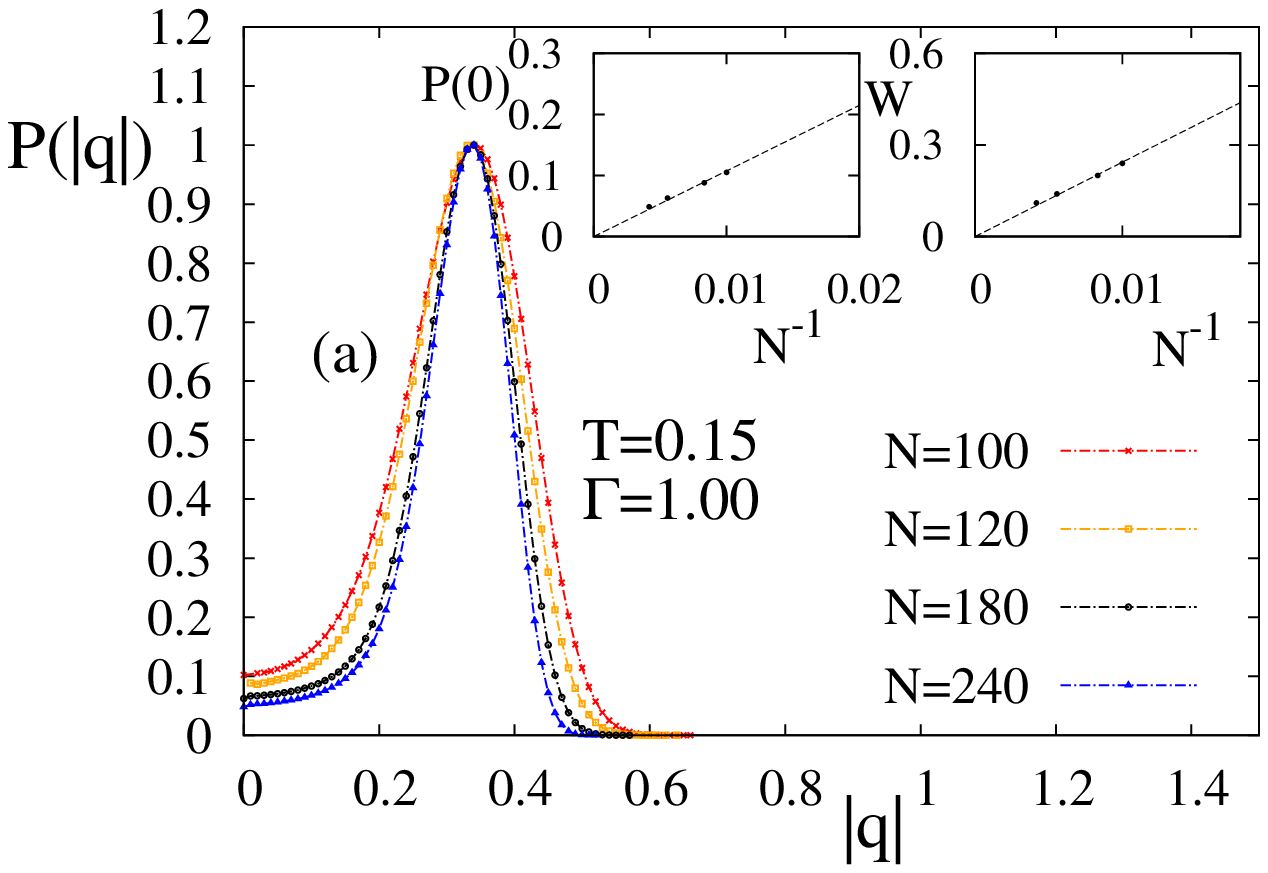}
\includegraphics[width=5.5cm]{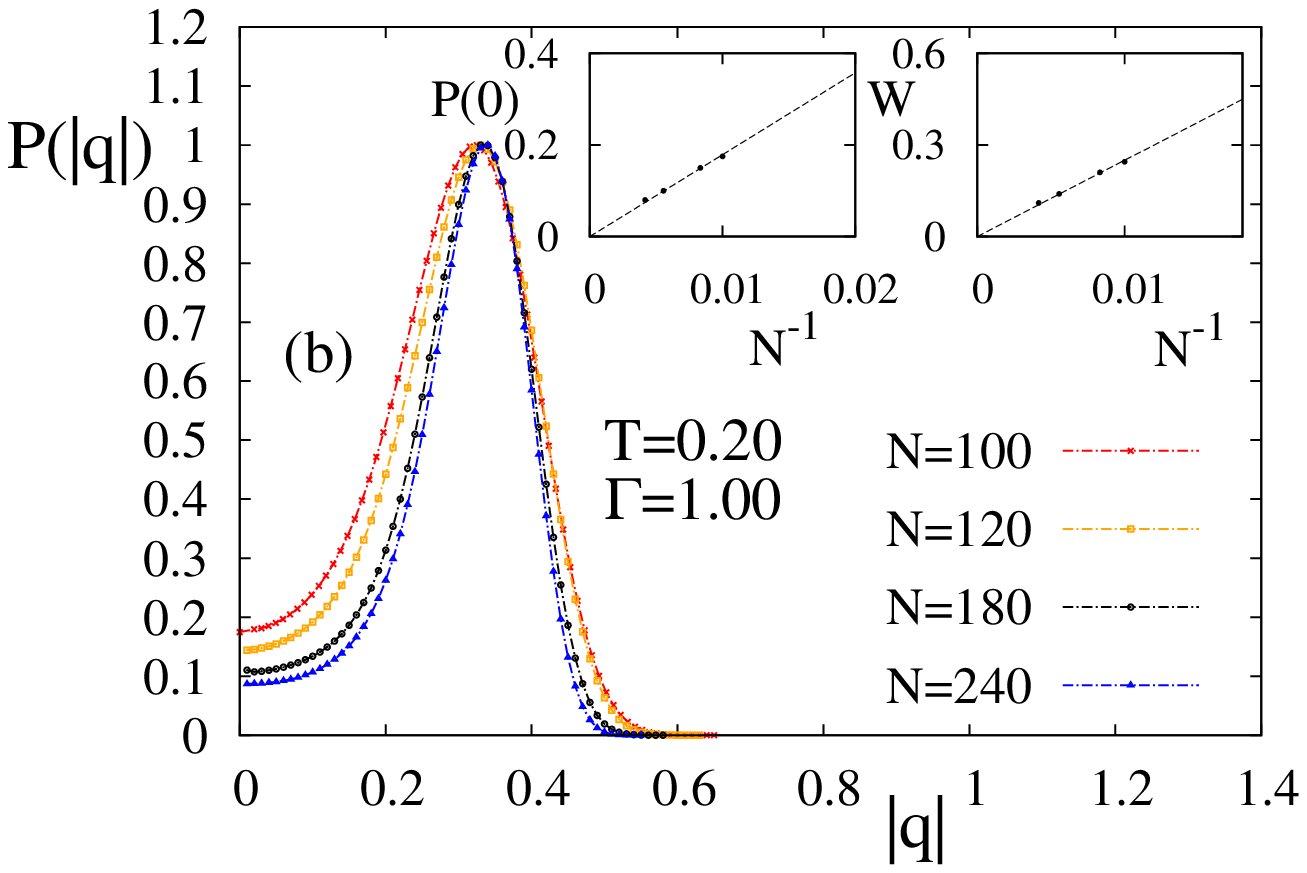}
\end{center}
\caption{(Color online) 
Plots of the peak-normalized order parameter distribution $P(|q|)$ for given sets of transverse field 
$\Gamma$ and temperature $T$, obtained from Monte Carlo simulations: (a) for $T=0.15$ and $\Gamma = 1.00$, 
(b) for $T = 0.20$ and $\Gamma = 1.00$. Extrapolations of both $P(0)$ and $W$ as functions of $1/N$ are shown in the insets.
In both cases the extrapolated values of these quantities go to zero in the infinite-system-size limit.}
\label{ergodic}
\end{figure*}
\begin{figure*}[htb]
\begin{center}
\includegraphics[width=5.5cm]{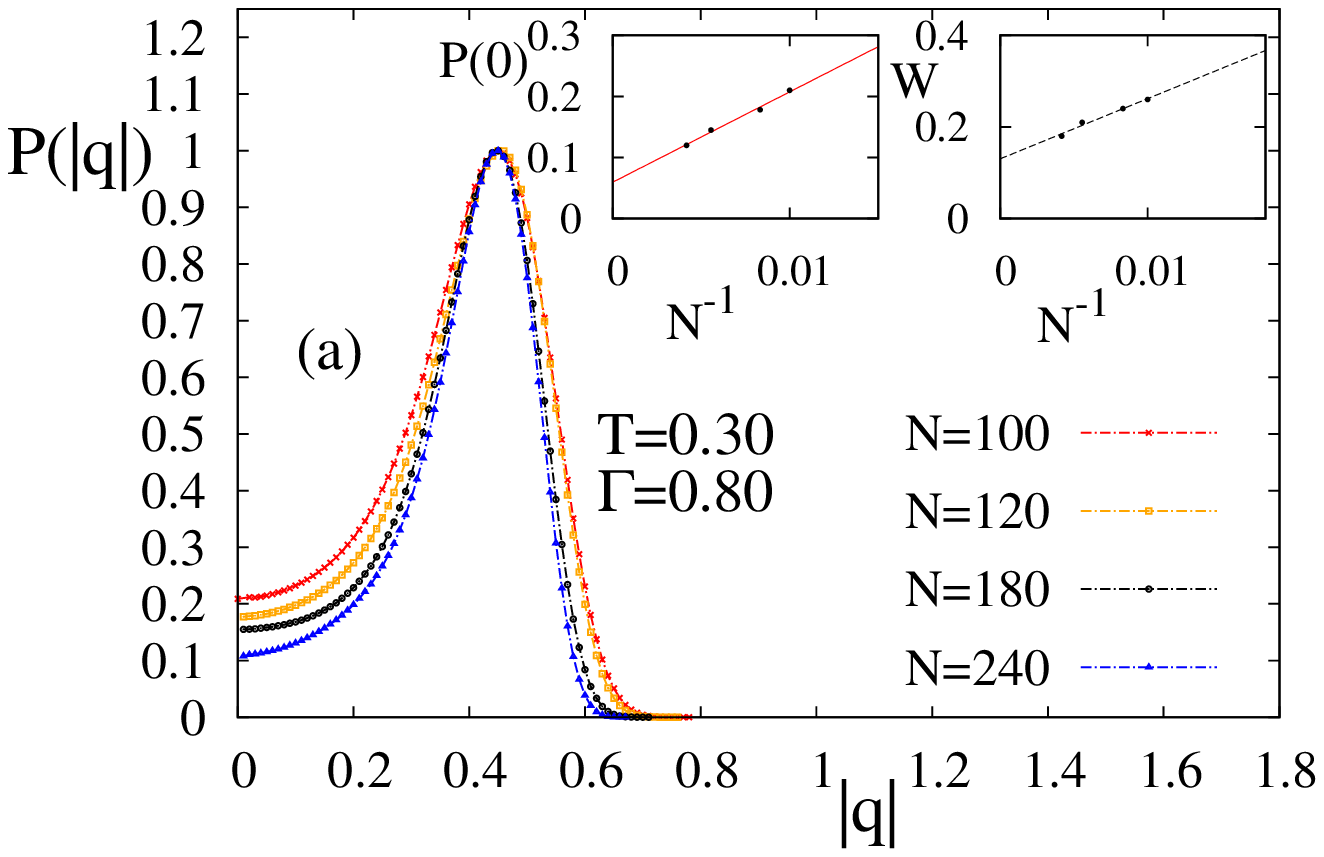}
\includegraphics[width=5.5cm]{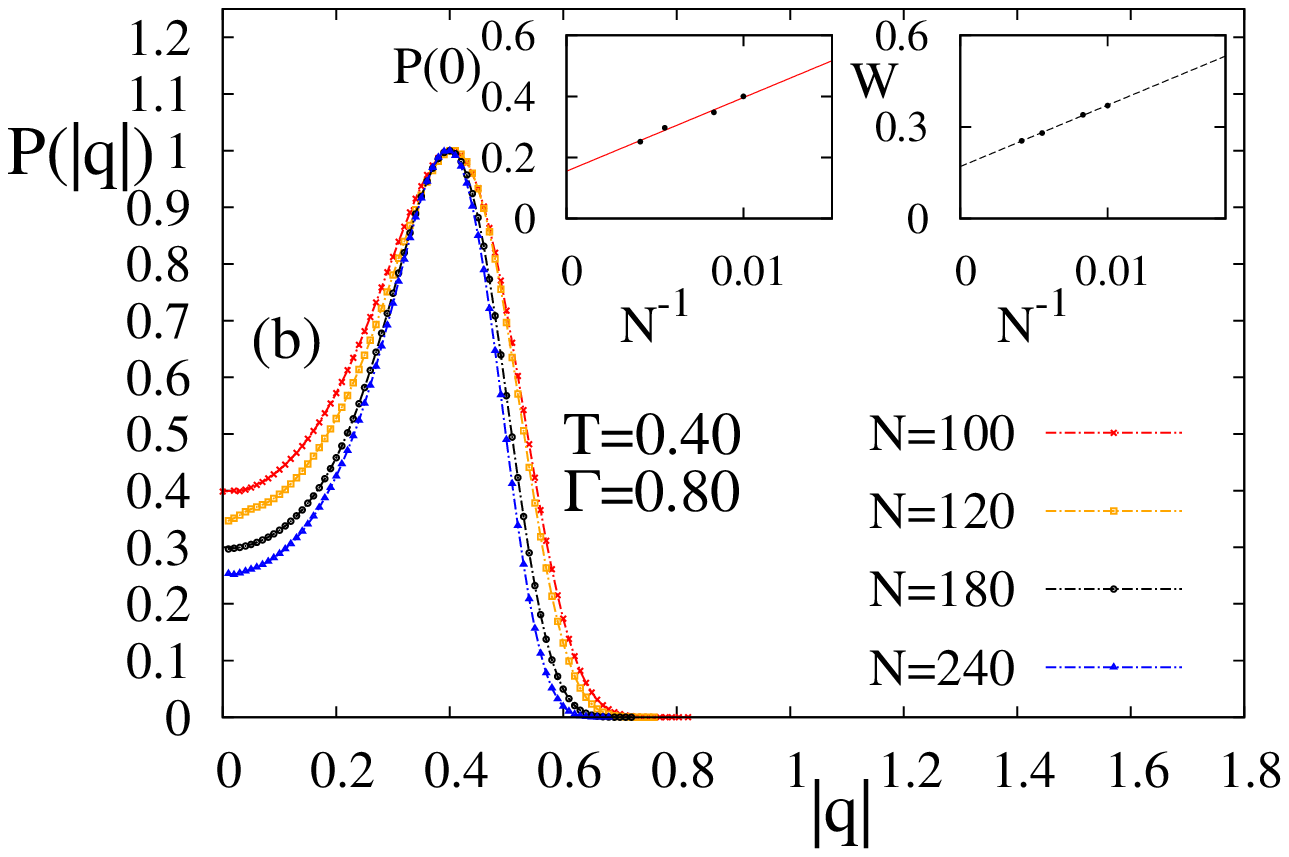}
\end{center}
\caption{(Color online) 
Plots of the peak-normalized order parameter distribution $P(|q|)$ for given sets of transverse field 
$\Gamma$ and temperature $T$, obtained from Monte Carlo simulations: (a) for $T=0.30$ and $\Gamma = 0.80$, 
(b) for $T = 0.40$ and $\Gamma = 0.80$. Again the insets show the extrapolations of $P(0)$ and $W$ as functions of $1/N$. 
In these cases the extrapolated values of $P(0)$ as well as $W$ remain finite even in the thermodynamic limit.}
\label{non-ergodic}
\end{figure*}
\subsection{Results of Monte Carlo simulations}
To find the order parameter distribution in the spin glass phase at finite temperatures, we perform Monte 
Carlo simulations on the effective classical Hamiltonian $H_{eff}$. We define the order parameter $q$ of 
the system as $q = \frac{1}{MN}\sum_{m=1}^{M}\sum_{i=1}^{N}\overline{{\langle \sigma_i^m\rangle}^2}$. 
Therefore, the order parameter distribution $P(q)$ can be evaluated as 
$P(q)=\overline{ \frac{1}{t_1}\sum_{t=t_0}^{t_0+t_1}\delta(q-q^{\alpha \beta}(t)) }$. For a given
set of $T$ and $\Gamma$ we compute both area-normalized and peak-normalized order parameter distributions. 
In the case of peak normalization the distribution is normalized by its maximum value.

For the finite temperature study, we perform Monte Carlo simulations with system sizes $N = 100, 120, 180, 240$ 
and $M=15$ Trotter slices. We notice that the equilibrium time  of the system is not 
uniform throughout the entire $\Gamma - T$ plane. Within the region $T < 0.25$ and $\Gamma < 0.40$ 
the system (for $100 \leq N \leq 240$) typically takes 
 $\lesssim 10^6$ time steps for equilibration, whereas the equilibrium time becomes $\lesssim 10^5$  
for the rest of the spin glass phase region. The thermal average is taken over $t_1=1.5 \times 10^{5}$ time 
steps and we take $1000$ samples for disorder averaging. As the system has ${\mathbb{Z}}_2$ symmetry we evaluate 
the distribution of $|q|$ instead of $q$. We notice a system size dependence of the value of $P(0)$. To 
find the value of $P(0)$ in the thermodynamic limit we extrapolate it as a function of $1/N$. In addition to the finite-size 
scaling of $P(0)$, we also estimate the value of $W$ for an infinite system size. Here $W$ is the width 
of the distribution function and is defined as $W = |q_2 - q_1|$. The distribution function becomes half 
of its maximum at $q = q_1$ and $q_2$. In the spin glass phase we find two distinct natures of the extrapolated 
values of both $P(0)$ and $W$. At low-temperature (high-transverse-field) the values of both $P(0)$ and $W$ 
tend to zero as the system size goes to infinity [see Fig.~\ref{area_ergodic}(a)]. This observation indicates
that in the thermodynamic limit $P(|q|)$ approaches the Gaussian form, which essentially suggests the ergodic 
behavior of the system. In contrast to this scenario, we also find a region (high $T$ and low $\Gamma$) in the
spin glass phase where neither  $P(0)$ nor $W$ vanishes even in the thermodynamic limit [see Fig.~\ref{area_ergodic}(b)]. 
There seems to be no possibility of $P(|q|)$ approaching a distribution with the Gaussian form in the large-system-size limit. 
Such behavior of $P(|q|)$ indicates that the system is nonergodic in this region of the spin glass phase. 
For more accurate measures of the ergodic and nonergodic regions in the spin glass phase, we also extract the 
behavior of the peak-normalized order parameter distribution. Again we find that $P(0)$ and $W$ of the 
peak-normalized distribution go to zero in the thermodynamic limit in this region of the spin glass phase, which has already 
been identified as the ergodic region from the study of the area-normalized distribution. This feature of the peak normalized 
distribution is shown in Figs.~\ref{ergodic}(a) and \ref{ergodic}(b). Similarly to the area-normalized distribution, at high temperature 
and low transverse field the values of $P(0)$ and $W$ for the peak-normalized distribution remain finite even in the 
large-system-size limit [see Figs.~\ref{non-ergodic}(a) and \ref{non-ergodic}(b)].

\begin{figure*}[ht]
\begin{center}
\includegraphics[width=5.5cm]{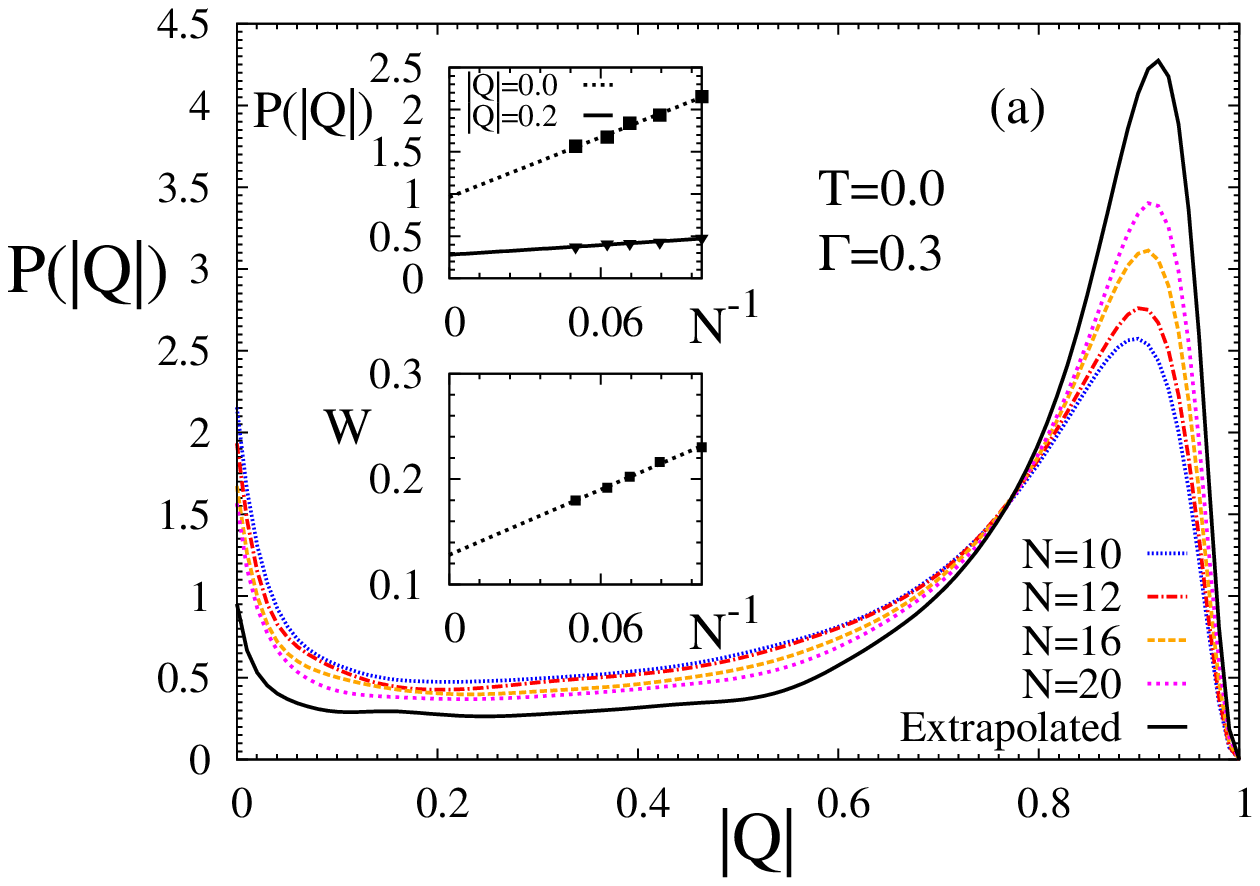}
\includegraphics[width=5.5cm]{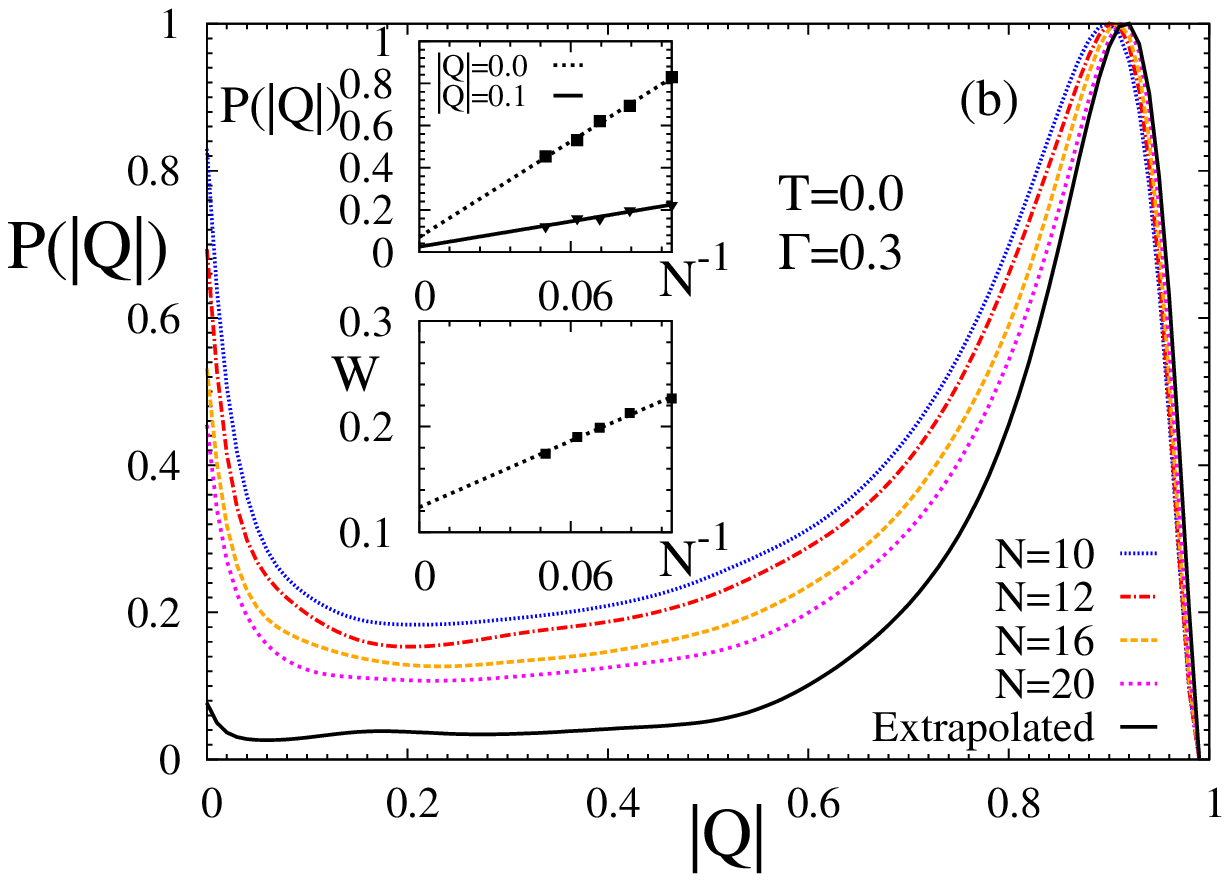}
\end{center}
\caption{(Color online) 
Variation of $P(|Q|)$ as a function of $|Q|$ for quantum SK spin glass with four different 
system sizes at $T=0$ and $\Gamma=0.3$. The numerical results are obtained using the exact 
diagonalization method. For (a) {the area of} each $P(|Q|)$ curve for a given value of $N$ is normalized to unity, 
whereas for (b) the peak of each $P(|Q|)$ curve is normalized by its maximum value. The 
top insets show the typical extrapolations of $P(|Q|)$ as functions of $1/N$ for (a) $|Q|=0.0$ and $0.2$, 
(b) $|Q|=0.0$ and $0.1$. The bottom 
inset of each figure shows the extrapolation of $W$ as a function of $1/N$.  
 }
\label{op_dist_0.3gama}
\end{figure*}

\subsection{Results of zero temperature exact diagonalizations}    
We use the exact diagonalization technique to study the nature of the order parameter distribution at zero temperature. 
The exact diagonalization of the quantum spin glass Hamiltonian $H$ [Eq.~(\ref{Ham})] is carried out by the Lanczos algorithm. 
Using this algorithm we evaluate the ground state of the system up to the system size $N = 20$. At zero temperature the 
order parameter of the system is defined as $Q = (1/N) \sum_i \overline{\langle\psi_0|\sigma_i^z|\psi_0\rangle^2}=(1/N)\sum_i\overline{Q_i}$. 
Here $Q_i$ denotes the site-dependent local order parameter, and the corresponding distribution of the  
local order parameter is given by $P(|Q|)=\overline{\frac{1}{N}\sum_{i=1}^N\delta(|Q|-Q_i)}.$
We numerically calculate $P(|Q|)$ for the system sizes $N = 10, 12, 16, 20$, which are very small. We study 
the behaviors of $P(|Q|)$ for several values of $\Gamma$ (at $T = 0$) in the spin glass phase. Similarly to the finite-temperature 
analysis, we investigate both the area- and peak-normalized $P(|Q|)$ [see Figs.~\ref{op_dist_0.3gama}(a) and \ref{op_dist_0.3gama}(b)]. 
When the system is in the spin glass state, $P(|Q|)$ shows a peak at a finite value of $|Q|$ along with a nonzero weight at $Q=0$. 
Although one can find an upward rise of $P(|Q|)$ as $|Q| \to 0$, the value of $P(0)$ decreases with increasing system size. In order to find the nature 
of both the area- and peak-normalized $P(|Q|)$ in the thermodynamic limit, we extrapolate $P(|Q|)$ as a function of $1/N$ for each values of 
$|Q|$. The extrapolations of $P(|Q|)$ at the values $|Q|=0$ and $0.1$ [for Fig.~\ref{op_dist_0.3gama}(a)] and $|Q| = 0$ and $0.2$ [for Fig.~\ref{op_dist_0.3gama}(b)] are shown in the top insets.
We also study the finite-size scaling of $W=|Q_2-Q_1|$, where  the $P(|Q|)$ becomes half of its maximum value at $Q_2$ 
and $Q_1$. We extrapolate $W$ as a function of $1/N$ to find its value in the large-system-size limit [see the bottom insets of 
Figs.~\ref{op_dist_0.3gama}(a) and \ref{op_dist_0.3gama}(b)]. Although due to the severe limitation of the maximum system size, 
the extrapolated curve does not take a delta-function-like shape, the distribution clearly becomes narrower with the increase in 
system sizes. The limitation in the system size also be the reason for observing a nonzero value of $W$ even in the thermodynamic 
limit. However, we infer that at zero temperature for any finite value of $\Gamma$, the $P(|Q|)$ curve will eventually become 
a delta function at a finite value of $|Q|$ in the thermodynamic limit.

\begin{figure}[h]
\begin{center}
\includegraphics[width=6.3cm]{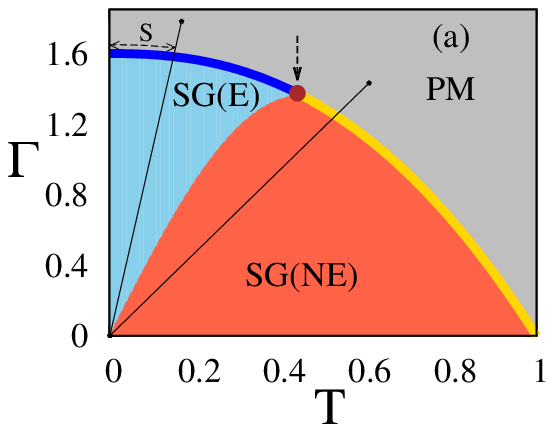}
\includegraphics[width=6.8cm]{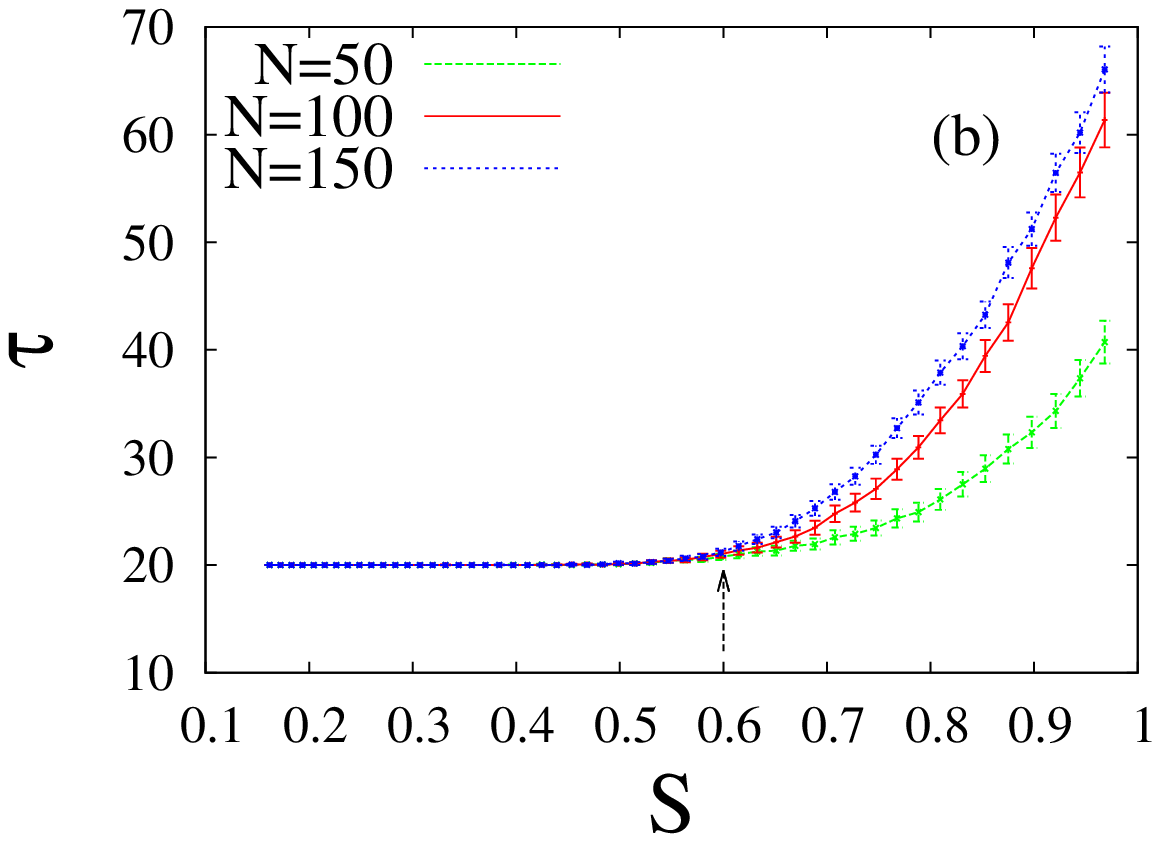}
\end{center}
\caption{(Color online)  
(a) Schematic phase diagram of the quantum SK model \cite{sudip-cl_qm}. The spin glass 
and paramagnetic phases are denoted by SG and PM, respectively. Our numerical results show that, in the case of 
ergodicity, the spin glass phase is further divided into two regions. They are the ergodic spin phase SG(E) and nonergodic 
spin glass phase SG(NE). The quantum-classical crossover point in the critical behavior of the model~\cite{sudip-cl_qm,sudip-yao} is 
indicated by the red dot on the SG-PM phase boundary. We perform annealing in the both SG(E) and SG(NE) regions along  
the linear paths, by simultaneously tuning  $T$ and $\Gamma$. Such annealing paths are indicated by the two inclined 
straight lines in the figure. (b) Variation of annealing time $\tau$ with $S$ (cf.~\cite{sudip-op_dis}). 
Here $S$ is the length of the arc calculated along the phase boundary starting from the zero-temperature quantum
transition point ($T=0, \Gamma \simeq 1.6$) and extending to the intersection of the annealing line with the 
phase boundary. The errors associated with the numerical data are indicated by the error bars. The annealing time does 
not have any system size dependence up to $S = 0.60 \pm 0.05$ (indicated in both figures by vertical arrows), which 
corresponds to $T = 0.49  \pm 0.03 , \Gamma = 1.31 \pm 0.04$. When the annealing paths pass through the SG(NE) 
region $\tau$ increases rapidly with the system size.
 }
\label{EE_NEE_phase_diagram}
\end{figure}

\section{Annealing through ergodic and nonergodic regions}\label{QA_EE}
Our investigations in the earlier sections clearly indicate the existence of a high-temperature (low-transverse-field) 
nonergodic region as well as a low-temperature (high-transverse-field) ergodic region in the spin glass phase. The line 
separating these two regions starts from $T = 0$, $\Gamma = 0$ and intersects the spin glass phase boundary 
at the quantum-classical crossover point~\cite{sudip-cl_qm,sudip-yao}. To find the dynamical features of these two regions, 
we study the annealing dynamics of the system through several paths using $H_{eff}$ with time-dependent $T$ and $\Gamma$. 
We vary the temperature and transverse field following the schedules 
$T(t) = T_0(1 - \frac{t}{\tau})$ and ${\Gamma}(t) = {\Gamma}_0(1 - \frac{t}{\tau})$, respectively. We choose $T_0$ and 
$\Gamma_0$ in such a way that the corresponding points on the phase diagram belong to the paramagnetic phase. In addition,  
they are equidistant from the critical line in the different parts of the phase diagram. We study the variation of the required 
annealing time of the system to achieve a very low free-energy associated with the very small values of $T \simeq 10^{-3} \simeq \Gamma$. 
At the end of the annealing schedule, we are forced to keep such small but nonzero values of the driving parameters to avoid the 
singularities in $H_{eff}$ and the annealing dynamics. We investigate the annealing of the system for a path that either passes 
through the quantum fluctuation dominated or classical fluctuation dominated [see Fig.~\ref{EE_NEE_phase_diagram}(a)] regions. 
Our numerical results show that when the annealing paths {pass} completely through the ergodic region, the annealing time becomes 
exclusively system-size-independent [see {Fig.~\ref{EE_NEE_phase_diagram}(b)}]. In contrast, for the paths that entirely lie in 
the nonergodic region, the annealing time increases monotonically with increasing $S$, a quantity measuring the 
arc-distance of the annealing line from the pure quantum ($T = 0$) transition point along the phase boundary [see Fig.~\ref{EE_NEE_phase_diagram}(b)]. 
We find that the numerical error in estimating the value of $\tau$, also increases monotonically with increasing $S$. 
In fact for $S \gtrsim 1.0$, the error bars in $\tau$ for different $N$ values start overlapping [see Fig.~\ref{EE_NEE_phase_diagram}(b)]. 
These results further confirm our earlier observation regarding the annealing time behavior reported in Ref.~\cite{sudip-op_dis}.

\begin{figure*}[htb]
\begin{center}
\includegraphics[width=6.0cm]{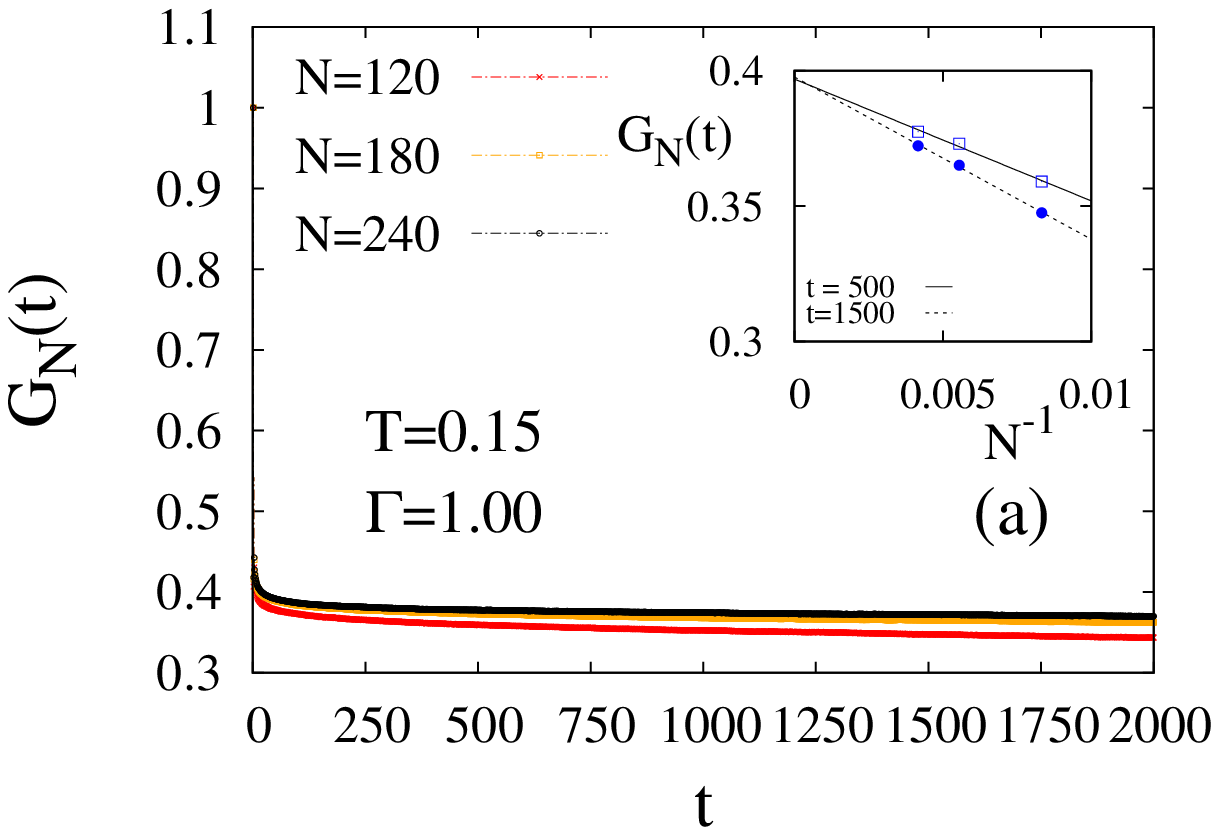}
\includegraphics[width=6.0cm]{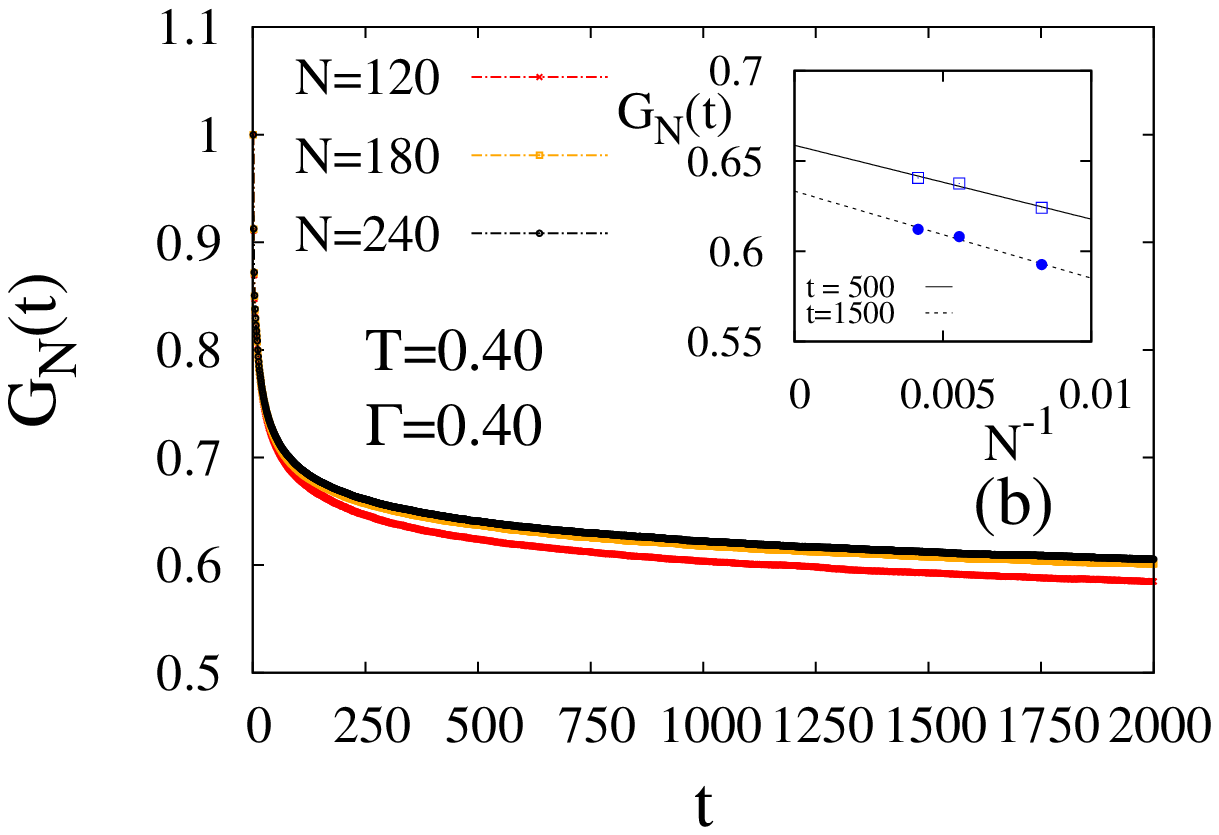}
\includegraphics[width=6.0cm]{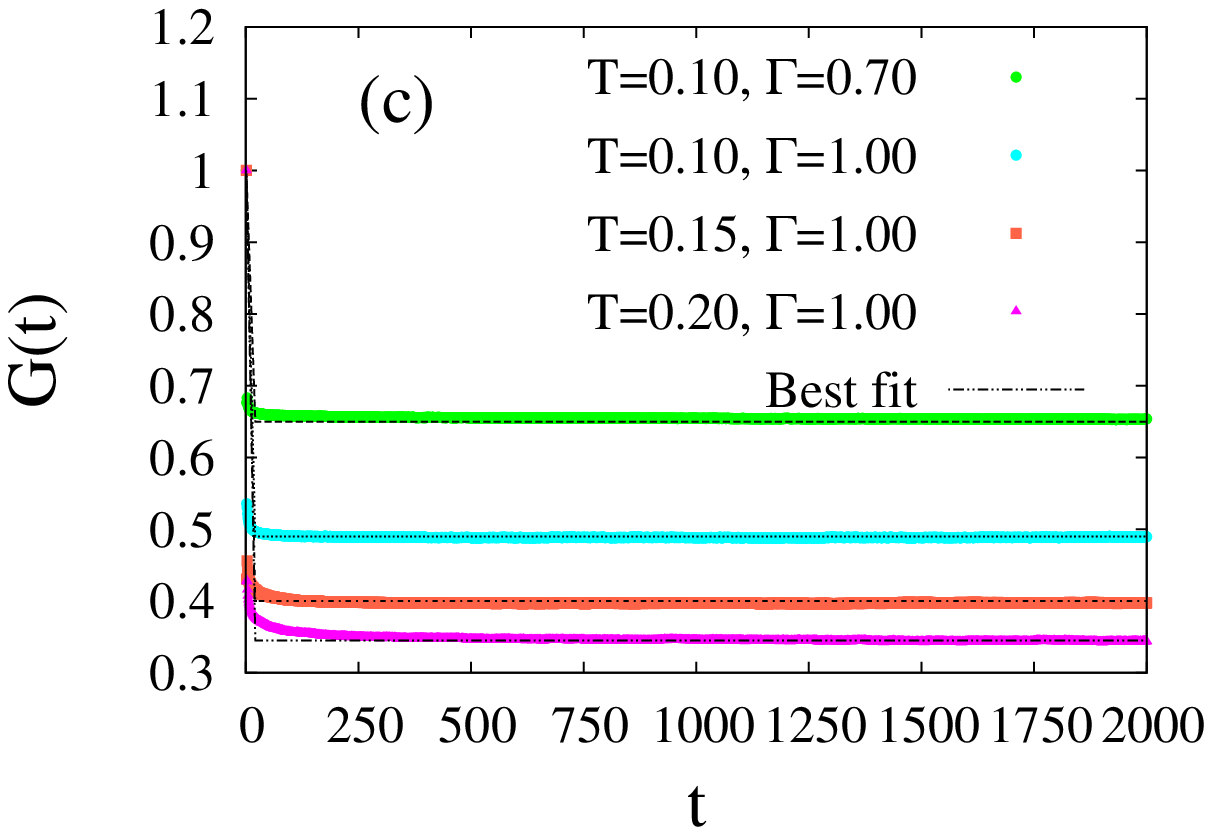}
\includegraphics[width=6.0cm]{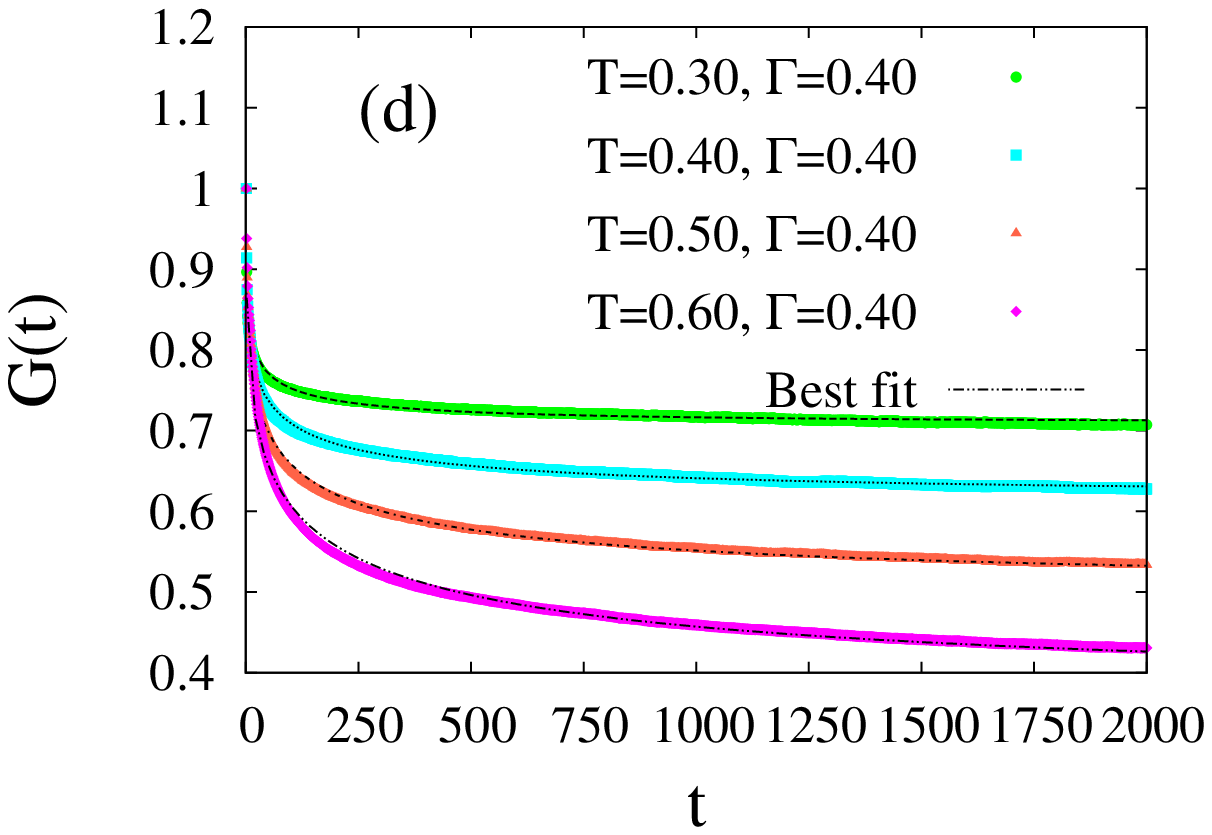}
\end{center}
\caption{(Color online) 
(a) Variation of spin autocorrelation function $G_N(t)$ [as defined in Eq.~(\ref{gnt})] with time $t$ for 
$T = 0.15$ and $\Gamma = 1.00$ with the system sizes $N = 120, 180, 240$. The inset shows the extrapolations of 
$G_N(t)$ a functions of $1/N$ at times $t = 500$ and $1500$. (b) Variation of autocorrelation $G_N(t)$ with identical 
system sizes at $T = 0.40$ and $\Gamma = 0.40$. Again the extrapolation of $G_N(t)$ at $t = 500$ and $1500$ are 
shown in the inset. (c) Variation of extrapolated autocorrelation $G(t)$ at ($T = 0.10, \Gamma = 0.70$), 
($T = 0.10, \Gamma = 1.00$),  ($T = 0.15, \Gamma = 1.00$), and ($T = 0.20, \Gamma = 1.00$). 
The best-fit [to Eq.~(\ref{auto-corr})] curves associated with these $G(t)$ variations are shown by the dotted lines. 
(d) Similar variations of $G(t)$ at ($T = 0.30, \Gamma = 0.40$), ($T = 0.40, \Gamma = 0.40$), ($T = 0.50, \Gamma = 0.50$),   
and ($T = 0.60, \Gamma = 0.40$) along with their corresponding best-fit lines.}
\label{auto_corr_plots}
\end{figure*}

\section{Study of spin autocorrelation dynamics}\label{corr}
We study the autocorrelation of the spins in both the ergodic and nonergodic regions of the spin glass phase. For fixed values of 
$\Gamma$ and $T$, after the equilibrium we consider a spin configuration (for a given disorder) at any particular Monte {Carlo} step $t_0$. 
Then we compute the instantaneous overlap of this spin configuration (at $t_0$) with the spin states pertaining to the consecutive 
Monte Carlo steps. We carry out this calculation for an interval of time $\mathbb{T}$, then with the spin profile at $\mathbb{T}+1$,
we repeat the same calculation for next $\mathbb{T}$ Monte Carlo steps. For a given system size $N$ the autocorrelation 
function is defined as
\begin{align}
 G_N(t) = \overline{\Big{\langle} \frac{1}{NM}\sum_{i=1}^N\sum_{n=1}^M\sigma_i^n(t_0)\sigma_i^n(t) \Big{\rangle}}. \label{gnt}
\end{align}
For each set of disorder we average $G_N(t)$ over several intervals, which is denoted by $\langle .. \rangle$. The disorder 
averaging is denoted by the overhead bar. Since we perform this calculation in the spin glass phase, the autocorrelation 
should decay to a finite value. We investigate the variation of $G_N(t)$ in both the ergodic and nonergodic regions and we notice a 
considerable difference in the relaxation behavior in these two regions. In the ergodic region the decay rate of the autocorrelation 
towards its equilibrium value is much faster than the decay rate in the nonergodic region. 

We perform Monte Carlo simulations with system sizes $N =120, 180, 240$ and $M = 10$ Trotter slices. The interval 
average is taken over $1000$ intervals and in each interval we consider $2000$ Monte Carlo steps. The disorder 
 average is taken over $100$ samples. The variation of $G_N(t)$ with $t$ for $T = 0.15$ and $\Gamma =1.00$ is shown 
in Fig.~\ref{auto_corr_plots}(a). We can see that the autocorrelation very quickly saturates (almost) to its equilibrium 
value. One can also see the system size dependence of $G_N(t)$. Therefore, we extract the autocorrelation $G(t)$ for an infinite 
system size through the extrapolation of $G_N(t)$ as a function of $1/N$. Such extrapolations at $t = 500, 1500$ are shown in the inset 
of Fig.~\ref{auto_corr_plots}(a). A similar plot of $G_N(t)$ for $T = 0.40$ and $\Gamma = 0.40$ (belonging to the nonergodic region) 
is shown in Fig.~\ref{auto_corr_plots}(b). One can clearly observe that in this case the decay of the autocorrelation is much slower than
 in the previous case. To estimate the relaxation time scale in the ergodic and nonergodic regions for an infinite 
system size, we try to fit the extrapolated curves $G(t)$ with the function 
\begin{align}
G(t) = G_s + (1 - G_s)\exp[-(\frac{t}{\tau_A})^{\alpha}]. \label{auto-corr}
\end{align}
Here $G_s$ is the tentative saturation value of $G(t)$ for the long-time limit and $\alpha$ is the stretched exponent. We refer to  
$\tau_A$ as the effective relaxation time of the system. The extrapolated curves $G(t)$ belong to the ergodic region and 
their corresponding best-fit lines are shown in Fig.~\ref{auto_corr_plots}(c). Since the fall of such $G(t)$ curves is 
extremely rapid, the fitting value of $\alpha$ is very high ($\approx 17 \pm 3$). The relaxation time $\tau_A$ in the ergodic 
region is typically on the order of $2$. The variations of $G(t)$ with $t$ in the nonergodic region with their associated 
best-fit lines are shown in Fig.~\ref{auto_corr_plots}(d). We find reasonably good fitting by considering $\alpha = 0.31 \pm 0.01$ 
but here we find that $\tau_A$ increases as we move deep into the nonergodic region from the line of separation between the ergodic 
and nonergodic regions. In Table~\ref{chart} we present the numerical results obtained from the fittings of $G(t)$ curves. 
From the numerical data we can clearly observe that similarly to critical exponent $\nu$, there is also a change in the value of the 
exponent $\alpha$ when we move from the ergodic to nonergodic region. Note, that the $G(t)$ variations 
of only four typical points in each of the SG(E) and SG(NE) regions are shown in Figs.~\ref{auto_corr_plots}(c) and \ref{auto_corr_plots}(d) 
[and analyzed with Eq.(7)]. Additional investigation for several other points in the regions also suggest similar conclusions.

\begin{table}[t]
\caption{Best-fit values of $G_s$, $\alpha$, and $\tau_A$ for different 
pairs of $T$ and $\Gamma$, where $G(t)$ is fitted to Eq.~(\ref{auto-corr}).}
\begin{center}
\begin{tabular}{|>{\small}c|>{\small}c|>{\small}c|>{\small}c|>{\small}c|}
 \hline
 {} & $T = 0.10$, $\Gamma = 1.00$ & $G_s = 0.49$ & $\alpha = 19.75$ & $\tau_A = 1.91$ \\ 
 {Ergodic} & $T = 0.15$, $\Gamma = 1.00$ & $G_s = 0.40$ & $\alpha = 16.64$ & $\tau_A = 1.87$ \\ 
 {(SG)} & $T = 0.20$, $\Gamma = 1.00$ & $G_s = 0.34$ & $\alpha = 14.34$ & $\tau_A = 1.90$ \\
 {} & $T = 0.10$, $\Gamma = 0.70$ & $G_s = 0.65$ & $\alpha = 13.67$ & $\tau_A = 1.86$ \\ \hline
 {} & $T = 0.30$, $\Gamma = 0.40$ & $G_s = 0.71$ & $\alpha = 0.30$ & $\tau_A = 11.01$ \\ 
 {Non-} & $T = 0.40$, $\Gamma = 0.40$ & $G_s = 0.62$ & $\alpha = 0.30$ & $\tau_A =  28.71$ \\ 
 {ergodic} & $T = 0.50$, $\Gamma = 0.40$ & $G_s = 0.51$ & $\alpha = 0.32$ & $\tau_A =  57.20$ \\ 
 {(SG)} & $T = 0.60$, $\Gamma = 0.40$ & $G_s = 0.38$ & $\alpha = 0.31$ & $\tau_A =  98.44$ \\ \hline 
\end{tabular}
\end{center}
\label{chart}
\end{table}

\section{Summary and discussion}
In sects.~\ref{critical_phenm}-\ref{QA_EE} we reviewed some of our earlier observations regarding 
the main question of our study here, and in sect.~\ref{corr} we reported our study of the autocorrelation 
behavior in the same model, confirming the earlier findings. 

We first discussed in sect.~\ref{critical_phenm} the determination of the phase diagram of the quantum SK model
(see Fig.~\ref{phase_diagram}) employing the Monte Carlo 
simulation (at finite temperatures) and exact diagonalization technique (at zero temperature). To extract the 
critical behavior at finite $T$, we considered system sizes $N = 20, 60, 180$ and chose the value of $M$ in 
accordance with the system size, keeping $M/N^{z/d}$ constant. At $T = 0$, we have a severe limitation of the system size (maximum $N = 22$). 
Here $d$ and $z$ respectively denote the effective dimension and dynamical exponent of the system. 
  We found that from the quantum transition point ($T = 0$, $\Gamma \simeq 1.63$) to almost the point 
($T = 0.45$, $\Gamma = 1.33$), the critical Binder cumulant ($g_c$)  
remains vanishingly small. Note that the critical Binder cumulant can effectively vanish even for
 (non-Gaussian) fluctuation-induced phase transitions~\cite{binder_92}. In this range of the phase boundary, 
we find the correlation length exponent $\nu \simeq 1/4$ from the data collapse of Binder cumulant plots. In the rest of 
 the phase boundary, the critical Binder cumulant is $g_c = 0.22 \pm 0.02$ and we observed a satisfactory data 
collapse with $\nu = 1/2$. These two different values of $g_c$ and $\nu$ for the two different parts of the phase 
boundary indicate the classical to quantum crossover (at $T \simeq 0.49$ and $\Gamma \simeq 1.31$) in the quantum SK model.

Unlike in the pure system, where the free-energy landscape is smoothly inclined towards the global minima, in the SK spin glass 
the landscape is extremely rugged. In particular, the local minima are often separated by macroscopically high energy barriers, 
inducing nonergodicity and a consequent replica-symmetry-broken distribution of the order parameter.  
Therefore, at any finite temperature the thermal 
fluctuation is unable to help the localized system to escape from the $O(N)$ free-energy barriers to reach the ground state (by 
flipping finite fraction of spins). With the aid of the transverse field the system can tunnel through such 
free-energy barriers~\cite{sudip-ray,sudip-das,sudip-bikas}. As a consequence, at low temperatures, the phase transition is governed by the quantum 
fluctuation and the system essentially exhibits  quantum critical behavior. 

We next studied (see sect.~\ref{op_distb}) the nature of the order parameter distribution $P(q)$ in the spin glass phase 
at finite temperatures through Monte Carlo simulations. For this numerical study we took $N = 100, 120, 180, 240$ and $M = 10$ 
[fixed; for small $N$ values, numerical results for $P(q)$ were found to remain fairly unchanged even when 
we varied $M$ with $N$ keeping $M/N^{z/d}$ constant]. We found 
[see Figs.~\ref{area_ergodic}(b),  \ref{non-ergodic}(a), and \ref{non-ergodic}(b)] that in the 
high-temperature (low-transverse-field) classical fluctuation dominated spin glass region,  along with the peak at 
the most probable value of the order parameter, the distribution contains a long tail (extending up to the zero 
value of the order parameter). This tail does not vanish even in the $1/N \to 0$ limit, which shows that the order parameter distribution 
remains Parisi type, corresponding to the nonergodic region SG(NE) [see Fig.~\ref{EE_NEE_phase_diagram}(a)] 
of the spin glass phase. On the other hand, we found [see Figs.~\ref{area_ergodic}(a), \ref{ergodic}(a), and \ref{ergodic}(b)] 
a low-temperature high-transverse-field region, where the 
order parameter distribution effectively converges to a Gaussian form (with a peak around the most probable value) 
in the infinite-system-size limit. This indicates the existence of a single (replica-symmetric) order parameter in this ergodic region SG(E) of the 
spin glass phase.  At zero temperature, we considered system sizes $N = 10, 12, 14,$ and $16$. 
Even with this limitation of the system size, the extrapolated order parameter distribution function showed [see Fig.~\ref{op_dist_0.3gama}(a)] a clear 
tendency to become one with a sharp peak (around the most probable value) in the large-system-size limit. 
We therefore conclude that the ergodic and nonergodic regions of the spin glass phase are separated by a line 
possibly originating from point ($T = 0, \Gamma = 0$) and extending up to the quantum-classical crossover point 
($T \simeq 0.49$, $\Gamma \simeq 1.31$)~\cite{sudip-cl_qm,sudip-yao} on the phase boundary [see Fig.~\ref{EE_NEE_phase_diagram}(a)]. 

To find the role of such quantum-fluctuation-induced ergodicity in the (annealing) dynamics, we investigated (see 
sect.~\ref{QA_EE}) the variation of the annealing time $\tau$ [required to reach close to the ground state(s)] 
with the system size following the schedules  $T(t) = T_0(1 - \frac{t}{\tau})$ and ${\Gamma}(t) = {\Gamma}_0(1 - \frac{t}{\tau})$. 
We attempted to reach a desired preassigned very low energy state (near the ground state) at the end of the annealing dynamics 
(in time $\tau$). We needed to keep both $T$ and $\Gamma$ nonzero (but very small) at the end of the annealing schedule  
as the Suzuki-Trotter Hamiltonian (which governs the annealing dynamics) has singularities at both $T = 0$ and 
$\Gamma = 0$. The values of $T_0$ and $\Gamma_0$ belong to the paramagnetic region of the phase diagram. We found 
[see Fig.~\ref{EE_NEE_phase_diagram}(b)] that the average annealing time does not depend on the system size when 
annealing is carried out along paths that pass through the ergodic region, whereas the annealing time becomes 
much larger and strongly size-dependent for paths that pass through the nonergodic region of the spin glass phase. 
These additional results, described in sect.~\ref{corr} confirm our earlier observations regarding the annealing time 
($\tau$) behavior reported in Ref.~\cite{sudip-op_dis}: Small values of $\tau$, independent of $N$, in the SG(E) region 
and order of magnitude larger $\tau$ values, growing with $N$, in the SG(NE) region. As indicated already in Ref.~\cite{sudip-ray}, 
all these phenomena are due to tunneling through macroscopically tall but thin free-energy barriers in the SK model. 

We performed another finite-temperature Monte Carlo dynamical study to distinguish the ergodic and nonergodic 
regions in the spin glass phase (see sect.~\ref{corr}). These results are newly reported in this paper. 
For given values of 
$T (> 0)$ and $\Gamma$, we investigated the temporal variation of the average spin autocorrelation $G_N(t)$ at finite 
temperatures by performing Monte Carlo simulations. 
We again considered system sizes $N = 120, 180, 240$ with $M = 10$. For each set of $T$ and $\Gamma$ values, using 
finite-size scaling of $G_N(t)$, we extracted the autocorrelation $G(t)$ for an infinite system size (see Fig.~\ref{auto_corr_plots}). 
The decay behavior of the extrapolated autocorrelation $G(t)$ 
is considerably different in the two regions. For the quantum-fluctuation-dominated spin glass region, the decay of $G(t)$ towards its 
equilibrium values is extremely fast. Our attempt to fit $G(t)$ with a stretched exponential function~[Eq.~(\ref{auto-corr})] gave the 
effective relaxation time $\tau_A \sim 2$ and a stretched exponent $\alpha$ of order $10$ 
(see Table~\ref{chart}; possibly indicating the failure of such a fit). 
 On the other hand, in the classical-fluctuation-dominated (nonergodic) region of the spin glass phase we obtained very good fits of the $G(t)$ curves  
with much larger values of $\tau_A$ and $\alpha = 0.31 \pm 0.01$ (see Table~\ref{chart}), again confirming the role of quantum tunneling. 
This observation of remarkably fast relaxation dynamics in the ergodic (quantum-fluctuation-dominated) region not only complements the findings~\cite{sudip-ray,sudip-cl_qm,sudip-op_dis} discussed in the earlier sections but also 
clearly indicates the origin of the success of quantum annealing~\cite{sudip_Kadowaki,sudip_Nishimori,sudip-das,sudip-ttc-book,sudip-lidar} through 
this region.


\acknowledgements
We are grateful to Arnab Chatterjee, Arnab Das, Sabyasachi Nag, Atanu Rajak, Purusattam Ray and Parongama Sen for their comments and suggestions. BKC gratefully acknowledges his J. C. Bose Fellowship (DST) Grant.


\end{document}